\documentclass[conference]{IEEEtran}
\usepackage{cite}
\usepackage{amssymb}
\usepackage{xcolor}

\usepackage{dblfloatfix}
\usepackage[normalem]{ulem}
\usepackage{amsfonts}
\usepackage{amsmath}
\usepackage{algorithmic}
\usepackage{graphicx}
\usepackage{textcomp}
\usepackage{hyperref}
\usepackage{cleveref}
\crefformat{section}{\S#2#1#3}
\crefformat{subsection}{\S#2#1#3}
\crefformat{subsubsection}{\S#2#1#3}
\usepackage{subcaption}
\usepackage{comment}
\usepackage{pifont}
\usepackage[cal=cm, calscaled=1.05]{mathalfa}
\usepackage{relsize}
\usepackage{cases}
\usepackage{tabularx}
\usepackage[utf8]{inputenc}
\usepackage{kotex}
\usepackage{multirow}
\usepackage{graphbox}
\usepackage{circledsteps}

\usepackage{setspace}
\usepackage{epstopdf}
\usepackage{makecell}
\usepackage[framemethod=tikz]{mdframed}
\usepackage[para]{threeparttable}
\usepackage{pythonhighlight}
\usepackage{colortbl}
\usepackage{mathtools}
\usepackage{array}
\usepackage{enumitem}
\usepackage[export]{adjustbox}

\newcommand{\suggestion}[1]{\bgroup\color{blue}#1\egroup\xspace}
\newcommand{\needtochange}[1]{\bgroup\color{red}#1\egroup\xspace}

\usepackage{kotex}
\usepackage[normalem]{ulem}
\usepackage{xcolor}
\usepackage{soul}
\usepackage{xspace}

\definecolor{mylime}{RGB}{205, 220, 57}
\definecolor{mygreen}{RGB}{60, 200, 0}

\definecolor{codegray}{rgb}{0.5,0.5,0.5}
\lstdefinestyle{pythonStyle}{
  columns=fullflexible,
  basicstyle=\linespread{0.95}\tiny\ttfamily\footnotesize,
  keywordstyle=\bfseries,
  morekeywords={return, NeighborApply, Pull, Apply},
  commentstyle=\color{codegray},
  frame=single,
  language=Python,
  stepnumber=1,
  numbers=left,
  numbersep=5pt,
  numberstyle=\scriptsize\color{codegray},
  tabsize=1,
  showspaces=false,
  showstringspaces=false,
  mathescape,
  moredelim=**[is][\color{purple}]{~}{~},
  moredelim=**[is][\color{blue}]{<}{>},
  moredelim=**[is][\color{orange}]{@}{@},
  literate={\\~}{{\textasciitilde}}1
  {\\<}{{\unichar{"003C}}}1
  {\\>}{{\unichar{"003E}}}1
  {\\@}{{\unichar{"0040}}}1
}

\newcommand{\mycomment}[1]{}
\usepackage[utf8]{inputenc}
\usepackage[T1]{fontenc}
\usepackage{soulutf8}
\usepackage{lipsum}
\usepackage[normalem]{ulem}

\makeatletter
\def\SOUL@hlpreamble{%
\setul{\dimexpr\dp\strutbox-2pt}{\dimexpr\ht\strutbox+\dp\strutbox-2pt\relax}
\let\SOUL@stcolor\SOUL@hlcolor
\SOUL@stpreamble
}
\makeatother

\newcommand\khlc[1][yellow]{
  \bgroup
  \markoverwith{\textcolor{#1}{\rule[-.5ex]{1pt}{2.5ex}}}
  \ULon
}
\newcommand{\todo}[1]{\bgroup\color{white}\textbf{\khlc[black]{TODO: [#1]}}\egroup\xspace}
\newcommand{\fixme}[1]{\bgroup\color{red}\textbf{\khlc{FIXME: [#1]}}\egroup\xspace}
\newcommand{\pointer}[1]{\bgroup\color{white}\textbf{\khlc[red]{POINTER: [#1 is working here]}}\egroup\xspace}
\newcommand{\reviewer}[1]{\bgroup\color{blue}#1\egroup\xspace}
\newcommand{\ranswer}[1]{\bgroup\color{red}#1\egroup\xspace}
\usepackage{pdfcomment}
\usepackage{marginnote}
\usepackage{calc}

\newlength{\markerHeight}
\newlength{\markerMargin}
\newlength{\linespace}
\newlength{\linedepth}

\sbox0{yf}
\tikzset{
    vertical align/.style={
        baseline=-.5*(height("$+$")-depth("$+$"))
    }
}
\usetikzlibrary{shapes.misc}
\tikzset{cross/.style={cross out, draw,
         minimum size=2*(#1-\pgflinewidth),
         inner sep=0pt, outer sep=0pt}}

\newcommand{\rotatetext}[1]{\rotatebox[origin=c]{90}{#1}}

\newcommand{\checkyes}{\tikz[vertical align]\draw[red,thick] (0,0) circle (.7ex);}
\newcommand{\checkno}{\tikz[vertical align]\draw[blue,thick] (-0.65ex,-0.65ex) -- (0.65ex,0.65ex) (-0.65ex,0.65ex) -- (0.65ex,-0.65ex);}
\newcommand{\checkneutral}{\tikz[baseline=.05ex] \draw[gray,thick] (0,0) -- (1.45ex,0) -- (0.725ex,1.25ex) -- cycle;}

\newcommand*{\squareNum}[1]{\tikz[baseline=(char.base)]{
            \node[shape=rectangle,fill,inner sep=1.8pt] (char) {\textcolor{white}{\scriptsize{#1}}};}}

\def\BibTeX{{\rm B\kern-.05em{\sc i\kern-.025em b}\kern-.08em
    T\kern-.1667em\lower.7ex\hbox{E}\kern-.125emX}}
\begin{document}

\title{\LARGE{GraphTensor: Comprehensive GNN-Acceleration Framework\\for Efficient Parallel Processing of Massive Datasets}\vspace{-10pt}}
\setstretch{0.925}

\author{
    \IEEEauthorblockN{Junhyeok Jang, Miryeong Kwon, Donghyun Gouk, Hanyeoreum Bae, Myoungsoo Jung}
    \IEEEauthorblockA{
        \textit{Computer Architecture and Memory Systems Laboratory}\\
        Korea Advanced Institute of Science and Technology (KAIST) \\
        http://camelab.org}
    \vspace{-25pt}
}

\maketitle

\begin{abstract}
    We present \textit{GraphTensor}, a comprehensive open-source framework that supports efficient parallel neural network processing on large graphs.
    GraphTensor offers a set of easy-to-use programming primitives that appreciate both graph and neural network execution behaviors from the beginning (graph sampling) to the end (dense data processing).
    Our framework runs diverse graph neural network (GNN) models in a destination-centric, feature-wise manner, which can significantly shorten training execution times in a GPU.
    In addition, GraphTensor rearranges multiple GNN kernels based on their system hyperparameters in a self-governing manner, thereby reducing the processing dimensionality and the latencies further.
    From the end-to-end execution viewpoint, GraphTensor significantly shortens the service-level GNN latency by applying pipeline parallelism for efficient graph dataset preprocessing. % being aware of GNN’s multi-layer architecture and data type dependency.
    Our evaluation shows that GraphTensor exhibits 1.4$\times$ better training performance than emerging GNN frameworks under the execution of large-scale, real-world graph workloads.
    For the end-to-end services, GraphTensor reduces training latencies of an advanced version of the GNN frameworks (optimized for multi-threaded graph sampling) by 2.4$\times$, on average.
\end{abstract}


%-------------------------------------------------------------------------------
\section{Introduction}
%-------------------------------------------------------------------------------
\label{sec:intro}
Graph neural networks (GNNs) are being paid significant attention and widely adopted in various computing systems such as
recommendation systems, social networks, and natural science \cite{ying2018graph,wang2019ngcf,song2019session,you2018graphmolecule}.
Typically, graph analyses and graph embeddings \cite{grover2016node2vec, lerer2019pytorch}
are considered time-consuming activities since they require processing all nodes of a target graph.
In contrast, GNNs only process a smaller number of nodes in the local graph connections to infer results for a given set of nodes by leveraging the learning process of convolutional neural networks (CNNs) \cite{hamilton2017inductive}. %\cite{hamilton2017inductive,graphsaint-iclr20,zou2019layer}.
GNNs can in turn make the graph-based analyses deliver ground-breaking performance and high interpretability \cite{kipf2017semi}.

However, their graph-natured data processing makes a difference between GNNs and the existing neural networks, such as CNNs and Transformers \cite{devlin2018bert}.
For example, GNNs aggregate multiple node feature vectors (i.e., \emph{embeddings}), which require traversing a target graph and processing a set of variables with sparse data.
To bridge the semantic gap, several studies extend programming abilities of \emph{deep learning} (DL) frameworks, such as TensorFlow and PyTorch.
Specifically, this type of GNN extension frameworks \cite{iclr19pyg, atc19neugraph, wang2021flexgraph, osdi21gnnadvisor} perform node aggregation and edge weight calculation by leveraging the underlying DL primitive operations and/or graph-oriented message passing interface \cite{wang2016gunrock, jia2017distributed}.
For the node aggregation, the extension accumulates all neighbor's embeddings to their destination vertex in parallel (edge-centric).
These edge-centric operations, unfortunately introduce lock and synchronization overhead when they update the embeddings at the destination, which significantly degrades performance \cite{malicevic2017everything}.
Hence, a few emerging GNN frameworks have recently applied vertex-centric data processing to GNN computing \cite{iclr19pyg, wang2019dgl, osdi21gnnadvisor, hu2020featgraph}.
Since the new version of GNN frameworks parallelizes the node accumulation across different vertices rather than edges, it can remove the synchronization issues and achieves higher bandwidth than before.

Despite these efforts, the emerging GNN frameworks yet suffer from low data processing performance on GNN computing due to three root causes.
First, the extension frameworks manage GNN tasks using existing DL operations or simulating graph processing without a complete understanding of GNN natures, which makes their vertex-centric processing not fully functional.
\emph{These partially vertex-centric operations make memory and cache management inefficient}, thereby limiting the scalability of parallel data processing in GPUs.
Specifically, leveraging DL operation approaches (e.g., \cite{iclr19pyg, osdi21gnnadvisor, atc19neugraph, wang2021flexgraph}) seriously waste the GPU internal memory to convert a sparse tensor to one or more dense tensors.
Similarly, simulating graph processing approaches (e.g., \cite{wang2019dgl, hu2020featgraph, MLSYS2020_gnnwithroc, husong2020g3}) dwindle the computation power because of their edge-wise thread scheduling algorithm.
The edge-wise thread scheduling in practice uses a parallel execution model optimized for conventional graphs, but not for GNNs \cite{wang2016gunrock, zhong2013medusa}. %This scheduling method spreads data across multiple GPU processors, which introduce cache bloat issue.
Second, the GNN extension frameworks aggregate node feature vectors first and then transform the aggregated feature vector using the multi-layer perceptron (MLP) in default.
\emph{This static kernel scheduling is unaware of the dimensionality reduction for node embeddings}, which are crucial to reducing the computation and memory requirement (thereby shortening the latency).
Lastly, \emph{existing GNN frameworks exhibit limited performance due to the long latency of GNN-specific preprocessing}, such as graph sampling, embedding lookup, and data transfers.
Such preprocessing is a per-service task and sits on the critical path in GNN computing, which makes the problem more significant.
We observe that the preprocessing latency for large-scale graphs accounts for 84.2\% of the total GNN processing time, on average.
We will closely analyze the aforementioned challenges in \cref{subsec:emerging_gnn_frameworks}.

In this paper, we propose a comprehensive open-source framework, \emph{GraphTensor}\footnote{All of the GraphTensor resources are available for free download on https://graphtensor.camelab.org} that supports parallel neural network processing on large graphs.
To directly expose the true benefits brought by diverse GNN models, GraphTensor offers a set of easy-to-use programming primitives that appreciate both graph and neural network execution behaviors from the beginning (graph sampling) to the end (dense data processing).
These primitives of GraphTensor can implement more than 315K different designs that can cover most architectural designs of GNNs \cite{you2020design}.
Overall, our comprehensive GNN acceleration framework consists of three research components: i) \emph{pure vertex-centric GNN computing}, ii) \emph{dynamic kernel placement}, and iii) \emph{end-to-end latency reduction}.

\noindent $\bullet$ \textbf{Pure vertex-centric GNN computing.}
GraphTensor traverses the input graphs and runs GNN kernels in a pure vertex-centric manner by considering the bounded, well-balanced edges of GNN input data.
Specifically, our programming primitives can directly aggregate/transform all the embeddings from exactly where GNN kernel threads look them up (destination-centric).
Such a method can remove the conversion from sparse tensor to dense tensor and minimize GPU memory usage.
Also, our primitives parallelize embedding processing through feature-wise thread scheduling rather than the existing edge-wise thread scheduling, which removes unnecessary global memory accesses.

\noindent $\bullet$ \textbf{Dynamic kernel placement.}
To overcome the static kernel placement (adopted by \cite{iclr19pyg, atc19neugraph, osdi21gnnadvisor}), an advanced version of GNN implementations requires manually modifying their kernel by being aware of GNN's hyperparameters before the execution \cite{wang2019dgl}.
Instead, GraphTensor automatically rearranges the GNN kernels to reduce the node dimensionality at the runtime, thereby shortening the latency of GPU computing further.
To achieve this, GraphTensor examines if GNN's feature vector transformation can cut down the dimension of feature vectors more than the aggregation.
Then, it conditionally performs the dynamic kernel placement at a construction time of GNN's dataflow graph (DFG).
We provide a simple cost estimation model to detect how much the node embeddings and corresponding computation can be reduced by appreciating all the hyperparameters, such as dimensions of input feature vectors, neighbor nodes, and hidden layers.

\noindent $\bullet$ \textbf{End-to-end latency reduction.}
Even though the two mechanisms above
can reduce the processing latency itself, the performance improvement of GNN computing is limited at the service-level due to the preprocessing overhead.
To take the overhead off the critical path in the GNN services, GraphTensor parallelizes and pipelines all the activities of GNN preprocessing by being aware of GNN's multi-layer architecture and data type dependency.
This pipeline parallelism can significantly shorten the preprocessing latency.
Specifically, it splits the preprocessing kernel into multiple meaningful subtasks and offers an execution chain, which can fully parallelize the subtasks.
It also considers different layers and data types of GNNs to minimize lock operations.

\begin{figure}
    \includegraphics[width=\linewidth]{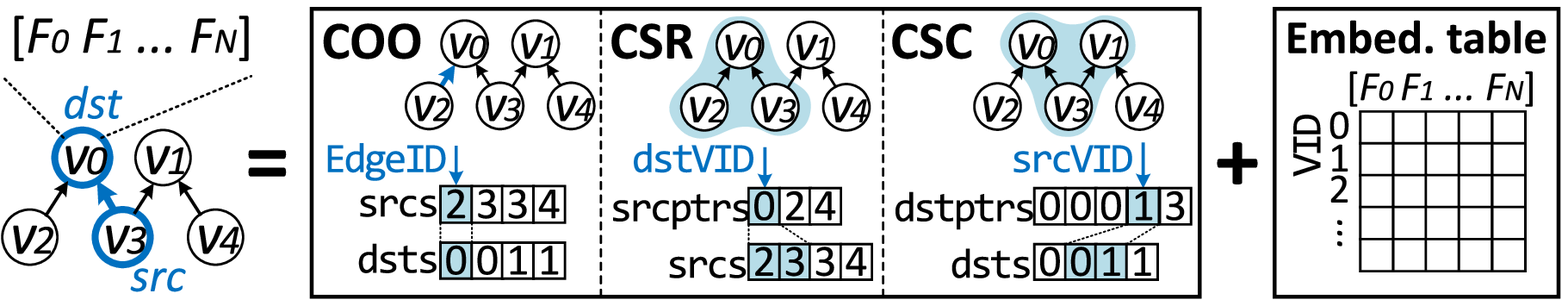}

    \vspace{-3pt}
    \begin{subfigure}{0.22\linewidth}
        \vspace{-18pt}
        \caption{Graph data.}
        \label{graph_data}
    \end{subfigure}
    \begin{subfigure}{0.53\linewidth}
        \vspace{-18pt}
        \caption{Graph storage formats.}
        \label{component_graph_structure}
    \end{subfigure}
    \begin{subfigure}{0.23\linewidth}
        \vspace{-18pt}
        \caption{Embedding.}
        \label{component_embedding}
    \end{subfigure}
    \vspace{-18pt}
    \caption{Components of graph data.}
    \vspace{-19pt}
    \label{graph_data_component}
\end{figure}

We implement GraphTensor based on TensorFlow, and evaluate diverse GNN models with a high-performance GPU.
Our evaluations show that GraphTensor exhibits 2.0$\times$ shorter training latency compared to the state-of-the-art GNN frameworks \cite{iclr19pyg, wang2019dgl,osdi21gnnadvisor}.
GraphTensor further shortens the service-level end-to-end latency of an advanced version (multi-threaded) of the GNN frameworks when processing a wide range of real-world graphs by 2.4$\times$, on average.

%-------------------------------------------------------------------------------
\section{Preliminaries}
%-------------------------------------------------------------------------------
\label{sec:background}

\subsection{Graph Neural Networks} \label{subsec:graph_neural_networks}
\textbf{Data representation.}
For graph processing in a GPU, a graph is often represented by an array-based sparse matrix and embedding table, each including the adjacency between nodes and node feature vectors (i.e., \emph{embeddings}), as shown in Figure \ref{graph_data_component}.
Based on different mechanisms to encode the adjacency, there are three representative graph storage formats: \emph{coordinate list} (COO), \emph{compressed sparse row} (CSR), and \emph{compressed sparse column} (CSC).
Figure \ref{component_graph_structure} depicts example graph structures of each format, all representing the same graph shown in Figure \ref{graph_data}.
COO represents a graph by using two arrays, each containing source (\emph{src}) and destination (\emph{dst}) vertex identifiers (\emph{VID}s).
By using edge ID as an index, we can easily retrieve the pair of src and dst VIDs of corresponding edge from the arrays.
While COO is an edge-centric storage format, CSR and CSC are designed toward vertex-centric processing.
They represent a graph using a vertex array that includes src VIDs (CSR) or dst VIDs (CSC).
CSR and CSC also have a pointer array of which element indicates the offset of the underlying vertex array of its edges.
The pointer array is indexed using a dst VID (CSR) or src VID (CSC).
On the other hand, the embeddings distill the vertex's information, which helps graph analysis, into a dense $n$-dimensional vector.
Typically, per-vertex embeddings are managed by contiguous memory, called \emph{embedding table} (Figure \ref{component_embedding}).
Note that the best storage format can vary based on how the target graph processing algorithm operates for GNNs.
In addition, since COO stores all pairs of adjacent vertices (edges), it exhibits heavier storage overhead than CSR/CSC, but COO can be easily translated to either CSR or CSC than the other formats. %\suggestion{COO에서 하나로 바꾸는 건 이득이지만, 두 포맷으로 모두 바꾸는 것은 오버헤드가 크다는 것 복선 깔기?}

\begin{figure}
    \vspace{-2pt}
    \includegraphics[width=\linewidth]{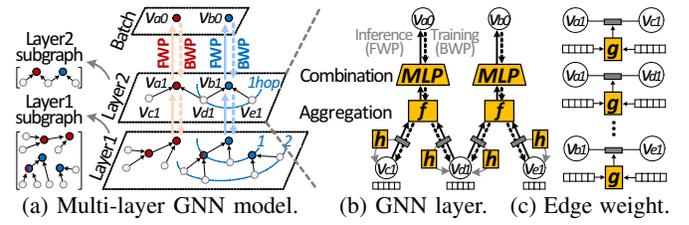}

    \vspace{-5pt}
    \begin{subfigure}{0.47\linewidth}
        \vspace{-15pt}
        \caption{Multi-layer GNN model.}
        \label{gnn_model_batch}
    \end{subfigure}
    \begin{subfigure}{0.27\linewidth}
        \vspace{-15pt}
        \caption{GNN layer.}
        \label{gnn_model_model}
    \end{subfigure}
    \begin{subfigure}{0.24\linewidth}
        \vspace{-15pt}
        \caption{Edge weight.}
        \label{gnn_model_edge_weight}
    \end{subfigure} \\
    \vspace{-17pt}
    \caption{GNN model overview.}
    \label{graph_model}
    \vspace{-18pt}
\end{figure}

\noindent \textbf{Algorithm model.}
As shown in Figure \ref{gnn_model_batch}, a vertex's output embedding can be speculated and/or trained by accumulating and transforming the input embeddings of its neighborhood \cite{kipf2017semi}.
When a GNN model accumulates and transforms the embeddings, called \emph{aggregation} and \emph{combination}, respectively,
the model traverses the vertices from neighbors in the outer-most loop to the target dst node, and it processes the corresponding embeddings for the vertices of each hop.
As a result, the model reflects all the input embeddings of the neighbors into the output embedding of the target dst node.
To this end, GNN receives an array of graphs (that includes vertices/edges within each hop) as input. GNN consists of multiple \emph{layers}, and each layer processes the corresponding hop's graph.
Figure \ref{gnn_model_model} shows how the layer 2 of a GNN processes vertices within 1 edge hop.
The GNN model's aggregation first finds out the src vertices of $V_{b1}$ (i.e., neighbors) and retrieves their embeddings from the embedding table.
It then accumulates the extracted embeddings using an aggregation function, $\pmb{f}$ (e.g., arithmetic mean).
The GNN model's combination then applies \emph{multi-layer perceptron} (MLP) to the accumulated embedding \cite{kipf2017semi}. %\cite{kipf2017semi, xu2018how}.
The MLP combines the different features of the accumulated embedding by using one or more linear transformations and non-linear functions, to achieve the output embedding representing $V_{b1}$ in a more accurate manner.
The combination's output is used for an input of the model's next layer computation.
This GNN processing terminates when there is no next layer to compute.
Note that multiple vertices can be given to the GNN model as a \emph{batch}.
Thus, the GNN model reiterates the aggregation and combination for all the remaining vertices in the batch ($V_{a1}$).

\begin{figure}
    \includegraphics[width=\linewidth,bb=0 6 507 77]{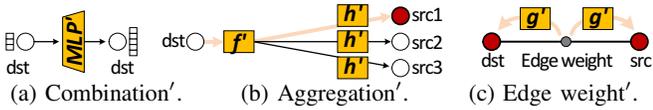}

    \vspace{-4pt}
    \begin{subfigure}{\linewidth}
        \centering
        \renewcommand*{\arraystretch}{0.3}
        \begin{tabularx}{\textwidth}{
                p{\dimexpr.31\linewidth-2\tabcolsep-1.3333\arrayrulewidth}% column 1
                p{\dimexpr.38\linewidth-2\tabcolsep-1.3333\arrayrulewidth}% column 2
                p{\dimexpr.31\linewidth-2\tabcolsep-1.3333\arrayrulewidth}% column 2
            }
            \vspace{-6pt} \caption{Combination$'$.}\label{training_combination}
             & \vspace{-6pt} \caption{Aggregation$'$.}\label{training_aggregation}
             & \vspace{-6pt} \caption{Edge weight$'$.}\label{training_edge_weight}
        \end{tabularx}
    \end{subfigure}
    \vspace{-19pt}
    \caption{GNN training operations.}
    \vspace{-19pt}
    \label{gnn_training}
    %\addtocounter{figure}{1}
\end{figure}

To consider the different meanings of each connection and the impacts of neighbor nodes, the GNN model can variously weight edges, which is referred to as \emph{edge weighting} (Figure \ref{gnn_model_edge_weight}).
The model applies an edge weight function, $\pmb{g}$, to all the src and dst vertices' embeddings (for each edge). %, \needtochange{including the target node's embedding}.
Since the model should reflect the weights to each edge appropriately, it employs an extra function $\pmb{h}$, that transforms the embedding of each edge's src node using $g$'s output vector.
Note that, to perform $g$ and $h$, this edge weighting needs to traverse the target graph, similar to what the aggregation performs.

\noindent \textbf{Training for GNNs.}
As shown in Figure \ref{gnn_model_batch}, an inference (forward propagation, \emph{FWP}) accumulates and transforms embeddings from the most outer loop to the target node(s). %in a given batch.
In addition to FWP, the training visits all the per-layer subgraphs and uses stochastic gradient descent by performing the GNN computation in the reverse order of the inference.
This backward propagation (\emph{BWP}) is aimed to calculate the derivative of error by FWP (loss) for weights in the target node.

While the BWP's derivatives of combination, aggregation, edge weighting are represented by the same functions of FWP,
an effective graph storage format for BWP can be different from that of FWP.
This is because the input and graph traversing order are different between BWP and FWP.
For example, as shown in Figure \ref{training_combination}, the derivative of the combination is equivalent to the combination using the transposed weight matrix,
which reverts the dst vertex's loss to the original embedding dimension space.
Similarly, $f'$ and $h'$ (i.e., the derivative of aggregation) are the same as the aggregation functions whose outputs are vectors for src vertices (not a dst vertex).
As shown in Figure \ref{training_aggregation}, this is because the input of $f'$ and $h'$ contains the loss information and will be propagated in reverse order.
Thus, while CSR fits well with FWP, CSC is better at traversing the graph in BWP.
Note that $g'$ is also equivalent to the FWP's weight computing kernel, but the derivative of the edge weight function is applied for both dst and src nodes (Figure \ref{training_edge_weight}).
This makes, in turn, edge-centric graph traversing better for edge weighting in FWP and BWP.

\subsection{Preprocessing for GNNs}
\label{subsec:preprocessing}
Inferring/training even a few vertices requires traversing a large region of graph nearby the target vertices and copying their embeddings from host memory to GPU memory.
The goal of preprocessing is reducing the number of vertices to compute and transfer the input graph between those two memories.
Since the preprocessing can significantly improve the performance of GNNs without a major loss of the model's accuracy \cite{hamilton2017inductive, ying2018graph},
it is key to accelerate GNN processing.
It is also crucial for scalability, as frameworks without preprocessing must store the entire graph in GPU memory.

\begin{figure}
    \includegraphics[width=\linewidth, bb=0 2 255 92]{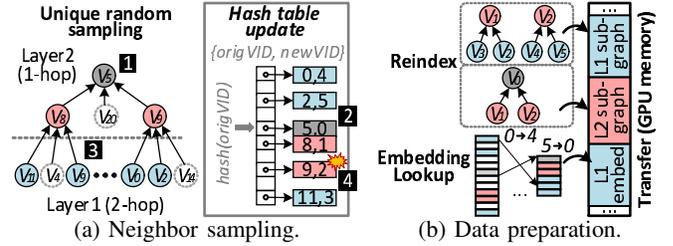}

    \vspace{-6pt}
    \begin{subfigure}{0.55\linewidth}
        \vspace{-10pt}
        \caption{Neighbor sampling.}
        \label{fig:pre_sample}
    \end{subfigure}
    \begin{subfigure}{0.43\linewidth}
        \vspace{-10pt}
        \caption{Data preparation.}
        \label{fig:pre_transfer}
    \end{subfigure}
    \vspace{-7pt}
    \caption{Illustration of GNN preprocessing.}
    \vspace{-20pt}
    \label{fig:gnn_preprocessing}
\end{figure}

\begin{figure*}
    \vspace{-5pt}
    \includegraphics[width=\linewidth]{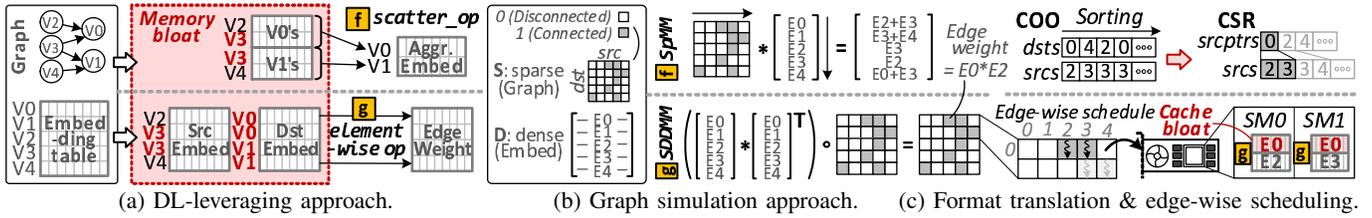}

    \vspace{-5pt}
    \begin{subfigure}{0.38\linewidth}
        \vspace{-24pt}
        \caption{DL-leveraging approach.}
        \label{bck_framework1}
    \end{subfigure}
    \begin{subfigure}{0.27\linewidth}
        \vspace{-24pt}
        \caption{Graph simulation approach.}
        \label{bck_framework2}
    \end{subfigure}
    \begin{subfigure}{0.34\linewidth}
        \vspace{-24pt}
        \caption{Format translation \& edge-wise scheduling.}
        \label{bck_framework3}
    \end{subfigure}

    \vspace{-6pt}
    \caption{Details of GNN extension framework.}
    \vspace{-18pt}
    \label{bck_framework}
\end{figure*}

\noindent \textbf{Neighbor sampling.}
Figure \ref{fig:gnn_preprocessing} explains necessary preprocessing components at the system-level GNN execution.
\emph{Neighbor sampling} samples the limited number of vertices from the graph by retrieving the adjacent nodes of dst vertices in a given batch and then picking $n$ vertices following a certain sampling priority (e.g., unique random \cite{hamilton2017inductive}). %(e.g. unique random \cite{hamilton2017inductive}, PageRank \cite{ying2018graph}).
In parallel, the sampling maintains a hash table for the sampled nodes.
The hash table allocates a new VID for each unique node added to the subgraph, starting from zero.
This new VID indexes a node among the other nodes in the small subgraph, whereas the original VID indexes a node in the large target graph.

Since each layer of GNNs processes neighbor nodes, ranging in different edge hops, the neighbor sampling prepares subgraphs for each of the GNN layers.
Figure \ref{fig:pre_sample} shows an example of the neighbor sampling.
This example employs a random sampling priority and assumes that $n$ is 2. %for the sake of brevity.
To prepare a subgraph for layer 2 (\emph{L2}), the sampling first selects two vertices ($V_8$, $V_9$) in the 1 edge hop from the dst vertex, $V_5$ (\squareNum{1}).
For the selected nodes, the sampling allocates new VID by inserting a pair of the new VID and original VID in a hash table (\squareNum{2}).
The sampling then randomly picks two neighboring vertices of $V_8$, in the two edge hops from $V_5$ for layer 1 (\squareNum{3}).
When sampling chooses a node already selected in previous steps (e.g., $V_9$), it does not allocate a new VID by scanning the hash table (\squareNum{4}).
It then iterates \squareNum{3} for all the previously sampled vertices ($V_9$) to complete the subgraph for layer 1.

\textbf{Data preparation.}
Figure \ref{fig:pre_transfer} shows the procedures of GNN data preparation, including graph reindexing, embedding lookup, and GPU transfers.
Each GNN layer execution in GPU frequently indexes a node in the input graph, which should be presented by one of the graph storage formats (COO/CSR/CSC).
To this end, a \emph{graph reindexing} algorithm renumbers the indices of subgraphs and prepares their graph structures, each corresponds to different GNN layer.
By referring the original VIDs in the sampled subgraphs, it retrieves newly allocated VIDs from the hash table and copies the VIDs to GPU's contiguous memory in the form of COO, CSR, or CSC.
This is equivalent to transferring each sampled subgraph to a set of distinct regions of the target GPU.
As discussed in \cref{subsec:graph_neural_networks}, it also requires preparing embeddings associated with the subgraphs.
Thus, \emph{embedding lookup} scans the global embedding table by using the original VIDs and allocates a new embedding table containing all embeddings of the sampled vertices.
Then, \emph{transfer} allocates memory space for the new table in GPU, and copies the new table from host to GPU memory.
This table is the one that GNN layer 1 needs to process.
GNN layer 2 also requires embeddings, but they will be the results of the layer 1's aggregation and combination.

Note that while the neighbor sampling and data preparation algorithms are vital for GNN acceleration,
they are a time-consuming task as they require traversing graphs, irregular scanning of several tables, and copying the corresponding data per service.

\section{Emerging GNN Frameworks} \label{subsec:emerging_gnn_frameworks}

There are several studies to explore new GNN programming interfaces and solutions \cite{iclr19pyg, wang2019dgl, atc19neugraph, osdi21gnnadvisor, MLSYS2020_gnnwithroc, hu2020featgraph, husong2020g3, wang2021flexgraph, rahman2020fusedmm},
which extend the existing deep learning (DL) frameworks such as TensorFlow and PyTorch through message passing interfaces.
Since most of them leverage a few successful public-frameworks \cite{abadal2020computing},
we classify the emerging GNN frameworks into two groups based on how they process the target datasets: i) DL-leveraging approach (\emph{DL-approach}) and ii) graph simulation approach (\emph{Graph-approach}). %, each simply being referred to as \emph{DL-approach} and \emph{Graph-approach}.
In general, DL-approach processes dense datasets for GNN kernels by revising the existing DL operations, while Graph-approach handles GNNs by directly processing the graph's sparse information.
In this section, we analyze the characteristics of the emerging GNN frameworks based on the two representative approaches. %and summarize them in Table \ref{tbl:prior_gnn_frameworks}.

\noindent \textbf{DL-leveraging approach.}
To reuse the underlying DL operations, DL-approach requires a \emph{sparse-to-dense data conversion} that collects the embeddings spreading across the embedding table (sparse) and composes a dense dataset as a matrix.
The top and bottom of Figure \ref{bck_framework1} illustrate the example operations of GNN aggregation and edge weighting, respectively.
For the aggregation, DL-approach extracts all neighbors' (src) embeddings from the embedding table and creates two input matrices for each dst vertices, $V_0$ and $V_1$.
Once it finishes sparse-to-dense data conversion, DL-approach computes $f$ using a DL operation/kernel (e.g., \texttt{scatter\_sum}, \texttt{scatter\_means}).
The DL kernel sums up the data of each column and composes an output row (per input matrix) of the result (aggregated) matrix.
The matrix's row and columns indicate the dst node and the corresponding embedding, respectively.
To weight the edges, DL-approach also performs the sparse-to-dense data conversion and creates two embedding matrices, for src and dst nodes, respectively.
It then implements $g$ using a DL operation such as (element-wise) \texttt{add} and \texttt{mul}, which generates an edge weight matrix by calculating each element of the two src and dst matrices;
the row and column of the result matrix contain the edge and corresponding weight vector, respectively.

\begin{figure}[b]
    \vspace{-18pt}

    \includegraphics[width=\linewidth]{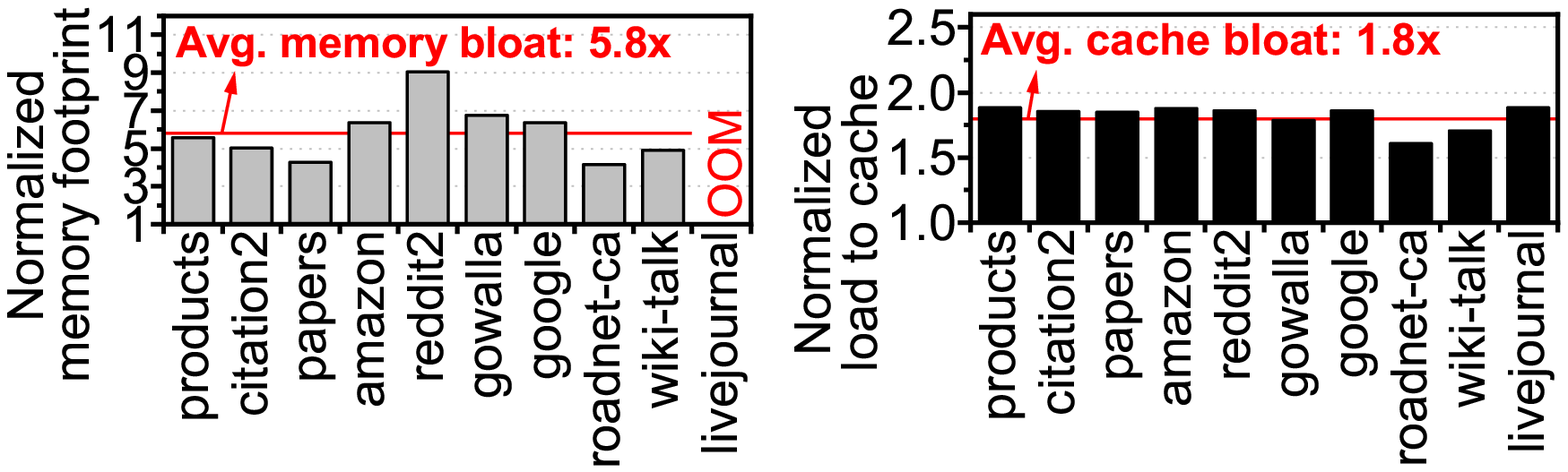}

    \hspace{0.07\linewidth}
    \begin{subfigure}{0.4\linewidth}
        \vspace{-8pt}
        \caption{Memory bloat.}
        \label{fig:chal_membloat}
    \end{subfigure}
    \hspace{0.1\linewidth}
    \begin{subfigure}{0.4\linewidth}
        \vspace{-8pt}
        \caption{Cache bloat.}
        \label{fig:chal_cachebloat}
    \end{subfigure}

    \vspace{-6pt}
    \caption{Challenges in GNN extension frameworks.}
    \vspace{-3pt}
\end{figure}

\noindent \textbf{Challenges of DL-approach.}
While the massive computing architecture of GPUs is well harmonized with the existing DL primitives, the sparse-to-dense conversion generates redundant embeddings in a GPU.
For example, in Figure \ref{bck_framework1}, the embeddings associated with the highlighted VIDs ($V_3$, $V_0$, and $V_1$) are redundantly stored in memory. % have many features all the same.
This GPU \emph{memory bloat} results in unnecessary data copies and significant waste of the GPU's internal memory.
Figure \ref{fig:chal_membloat} shows the memory footprint of DL-approaches.
To provdie an accurate representation, we normalized the value by the size of input embedding table.
The figure demonstrates that, on average, the memory bloat increases the the memory footprint by 5.8$\times$.
Recently, several DL apporach frameworks have addressed the memory bloat issue on aggregation \cite{osdi21gnnadvisor}.
However, they cannot avoid the issue on edge weight calculation that relies on DL operation-based user code.

\noindent \textbf{Graph simulation approach.}
While DL-approach uses DL operations only with the embeddings, Graph-approach employs sparse matrix multiplication-based operations working with both graph and embeddings.
Specifically, the graph dataset of Graph-approach (the left of Figure \ref{bck_framework2}) is represented by a sparse matrix ($S$), while its embedding table is considered as 2D dense matrix ($D$) with five embeddings ($E_0$-$E_4$).
The top and bottom of Figure \ref{bck_framework2} show how Graph-approach implements GNN aggregation and edge weighting with sparse matrix multiplication (\emph{SpMM}) and sampled dense-dense matrix multiplication (\emph{SDDMM}), respectively.

For the aggregation, SpMM can represent $f$ by multiplying sparse and dense matrices ($S*D$).
For example, $V_0$'s aggregation result is the same as the product of $S$'s first row and $D$'s first column ($E_2+E_3$).
On the other hand, SDDMM can implement the edge weighting ($g$) with multiplications of the dense matrix, its transpose and sparse matrix ($(D*D^T) \circ S$);
$D*D^T$ computes all edges' weight, supposing that all vertices are connected.
SDDMM then completes the edge weighting by filtering out the weight of disconnected edges using $S$.

Note that S is a conceptual matrix in Graph-approach.
Since it contains many `0's (disconnected), Graph-approach frameworks simulate S using COO or CSR for SpMM and SDDMM. %\needtochange{Specifically, for the SpMM, Graph approach uses src VIDs for each dst vertex to lookup and accumulate the src's embeddings. For the SDDMM, it uses the pairs of src, dst vertices for each edge to weight them using the pair of embeddings for src and dst.}
Specifically, for the SpMM, Graph approach uses src VIDs (for each dst vertex) to lookup and accumulate the src's embeddings. For the SDDMM, it uses the pairs of src, dst VIDs (for each edge) to retrieve their embeddings for weighting the corresponding edge.

\noindent \textbf{Challenges of Graph-approach.}
The simulation of SpMM and SDDMM can remove the memory bloat issue of the DL-approach as it does not need to prepare extra embeddings within a GPU.
However, Graph-approach also has two challenges: i) \emph{format translation} and ii) \emph{cache bloat}.
Graph-approach commonly uses COO since SDDMM needs edge information and COO exhibits lower overhead to convert into other storage format. %, except for ROC \cite{MLSYS2020_gnnwithroc} (section \ref{sec:relatedwork}).
However, as SpMM still needs src node information per dst vertex, it needs to translate COO to CSR.
As shown in the top of Figure \ref{bck_framework3}, Graph-approach sorts COO's src and dst arrays based on dst VIDs, and then it converts the dst array to src pointer array (srcptrs) to have CSR.
Note that, as explained in \cref{subsec:graph_neural_networks}, since BWP needs dst node information per src node, it also needs to translate COO to CSC.

On the other hand, the bottom of Figure \ref{bck_framework3} explains the \textit{cache bloat} issue that Graph-approach suffers from.
Similar to conventional graph frameworks \cite{wang2016gunrock, zhong2013medusa}, Graph-approach allocates a thread block per edge, and threads in the block process multiple features in the corresponding embedding (\emph{edge-wise scheduling}).
Note that an edge corresponds to a matrix element having value `1' in COO or CSR.
Since thread blocks are scheduled by different streaming processors (SMs) in parallel, different edges with the same dst are processed by different SMs.
Such scheduling places multiple copies of embeddings across the different SMs, introducing the cache bloat issue ($E_0$ in the figure).
Figure \ref{fig:chal_cachebloat} illustrates the significance of this issue for various graphs we tested.
We measured the cache data loaded from Graph-approach's SDDMM and normalized it by the size of the original embedding table.
One can observe from the figure that, cache bloat needlessly loads an average of 81.9\% more data to the cache,
making their cache management less efficient and increasing the global memory accesses, which can degrade overall performance.

%-------------------------------------------------------------------------------
\section{GraphTensor}
%-------------------------------------------------------------------------------
\label{sec:design-overview}
\begin{figure}
    \vspace{-4pt}
    \includegraphics[width=1\linewidth]{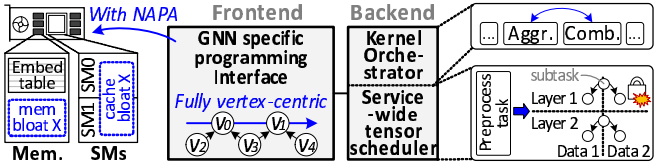}
    \vspace{-24pt}
    \caption{Overview of GraphTensor.}
    \vspace{-20pt}
    \label{fig:graphtensor_overview}
\end{figure}

\subsection{High-Level View.}\label{subsec:highlevel}
Figure \ref{fig:graphtensor_overview} shows the overview of GraphTensor supporting parallel neural network processing on large graphs.
In particular, it appreciates both graph processing and conventional deep learning characteristics and accelerates the entire GNN processing procedures of the target GPU as well as in the host,
thereby achieving high performance.

To this end, the frontend of our framework offers GNN-specific programming interfaces,
while its backend's kernel orchestrator and service-wide tensor scheduler ensemble the dataflow graph and parallelize GNN preprocessing.
Specifically, the frontend interfaces employ \emph{\textbf{\underline{N}}eighbor\textbf{\underline{A}}pply-\textbf{\underline{P}}ull-and-\textbf{\underline{A}}pply} (NAPA) programming model that treats edge weighting, aggregation,
and combination in a destination-centric manner; GraphTensor's destination-centric operations explore the graph around dst nodes (rather than src nodes) and parallelize embedding processing by focusing on dsts, not edges.
Our NAPA eliminates sparse-to-dense data conversion, which prevent GPU memory bloat and removes cache bloat over feature-wise thread scheduling.

Meanwhile, the backend's kernel orchestrator, rearranges the aggregation and combination for each GNN layer by taking into account the dimensionality of its input and output node embeddings.
This method, called \emph{dynamic kernel placement}, can reduce the amount of data to compute for GNN training. %by being aware of per-layer embedding dimensionality.
The other module, \emph{service-wide tensor scheduler}, splits the preprocessing task of GNNs into multiple subtasks and parallelizes their executions by considering data representation and computation dependency of node features and subgraphs.
In addition, the tensor scheduler addresses lock contentions, observed by parallel accesses of the hash table (used for neighbor sampling and graph reindexing).
We first explain the frontend in this section and then describe the backend modules in \cref{sec:implementation}.

\begin{figure}
    \includegraphics[width=1\linewidth, valign=t]{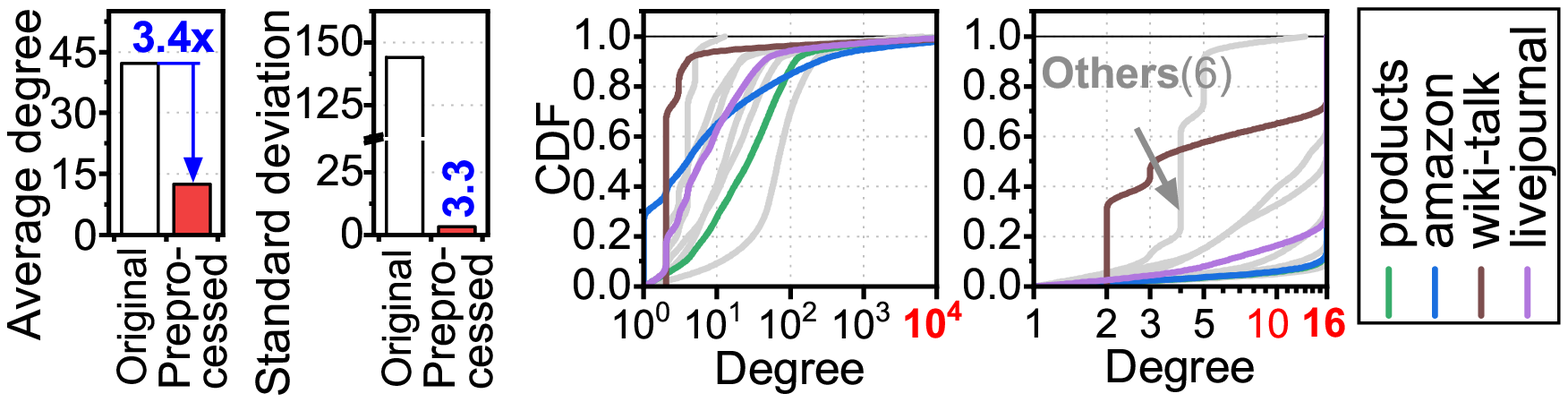}

    \begin{subfigure}[t]{0.33\linewidth}
        \vspace{-6pt}
        \caption{Degree statistics.}
        \label{fig:degree_stat}
    \end{subfigure}
    \hspace{0.02\linewidth}
    \begin{subfigure}[t]{0.21\linewidth}
        \vspace{-6pt}
        \caption{Original.}
        \label{fig:cdf_original}
    \end{subfigure}
    \begin{subfigure}[t]{0.3\linewidth}
        \vspace{-6pt}
        \caption{Preprocessed.}
        \label{fig:cdf_sampled}
    \end{subfigure}
    \hspace{0.03\linewidth}

    \vspace{-6pt}
    \caption{Degree distribution of graph.}
    \label{fig:sampled_graph_characteristics}
    \vspace{-18pt}
\end{figure}

\begin{figure}[b]
    \vspace{-20pt}
    \includegraphics[width=1\linewidth, valign=t]{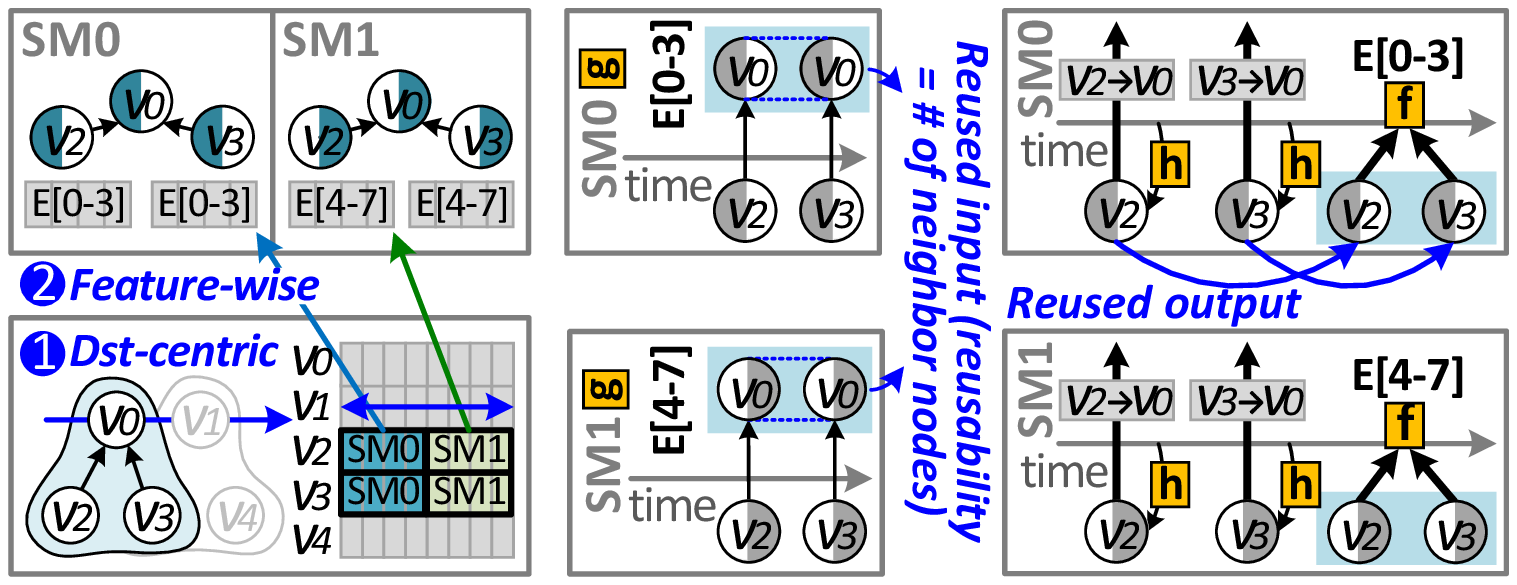}

    %\hspace{0.05\linewidth}
    \begin{subfigure}{0.42\linewidth}
        \vspace{-8pt}
        \caption{Thread scheduling.}
        \label{fig:pure_vertex_thread_sched}
    \end{subfigure}
    \begin{subfigure}{0.37\linewidth}
        \vspace{-8pt}
        \caption{NeighborApply.}
        \label{fig:pure_vertex_neighbor_apply}
    \end{subfigure}
    \begin{subfigure}{0.18\linewidth}
        \vspace{-8pt}
        \caption{Pull.}
        \label{fig:pure_vertex_pull}
    \end{subfigure}

    \vspace{-7pt}
    \caption{Dst-centric and feature-wise primitives.}
    \label{fig:over_pure}
    \vspace{-2pt}
\end{figure}

\begin{figure*}[b]
    \addtocounter{figure}{1}
    \vspace{-10pt}
    % h: 2.5
    % w: 3.2, 2, 3.92, 4.38
    \begin{minipage}[t]{0.61\linewidth}

        \vspace{-5pt}
        \begin{subfigure}[b]{.34\linewidth}
            \includegraphics[width=1\linewidth,bb=0 0 245 194]{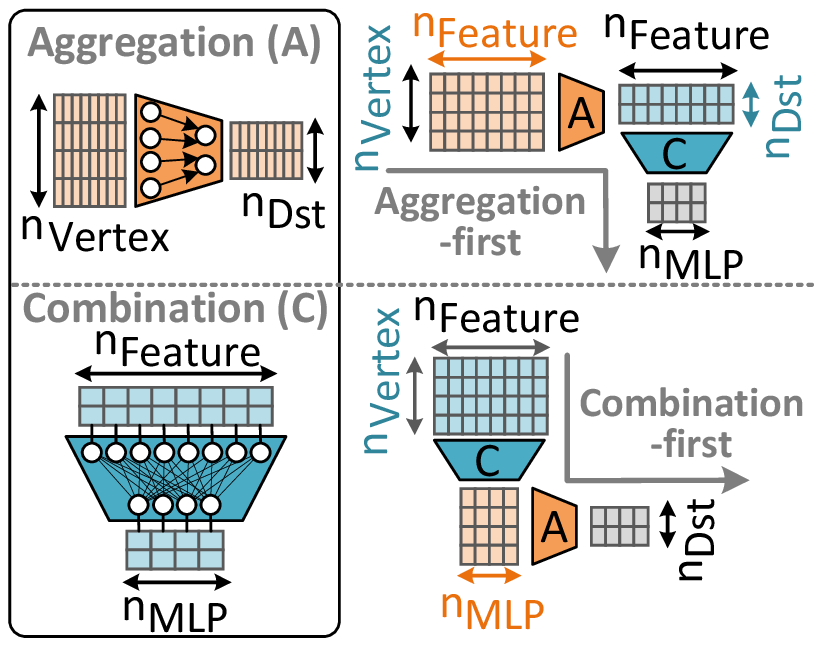}
            \vspace{-17pt}
            \caption{Dimensionality reduction.}
            \label{fig:dkp_concept}
        \end{subfigure}
        \begin{subfigure}[b]{.215\linewidth}
            \includegraphics[width=1\linewidth,bb=0 0 144 174]{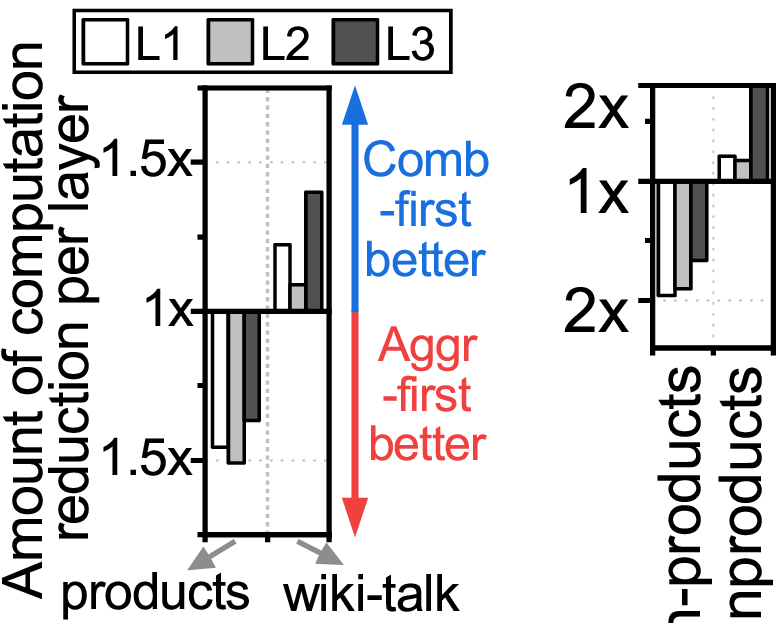}
            \vspace{-17pt}
            \caption{Motivation.}
            \label{fig:dkp_motiv}
        \end{subfigure}
        \begin{subfigure}[b]{.43\linewidth}
            \includegraphics[width=1\linewidth,bb=0 0 297 187]{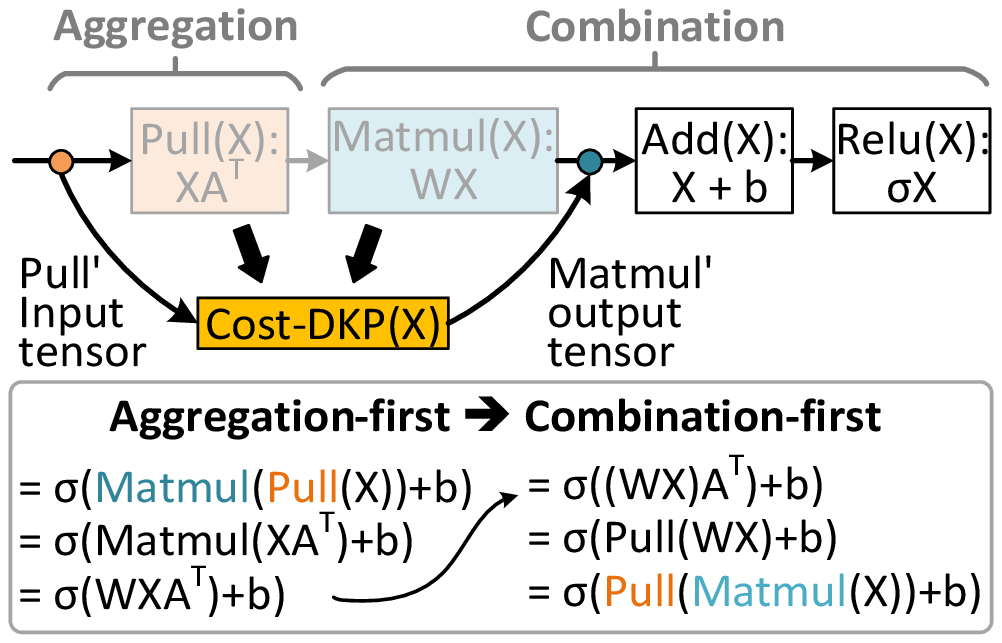}
            \vspace{-17pt}
            \caption{Dynamic kernel placement.}
            \label{fig:dkp_impl}
        \end{subfigure}
        \vspace{-19pt}
        \caption{Details of our GNN kernel orchestrator.}
        \label{design_dkp}
        \vspace{-20pt}
    \end{minipage}
    \begin{minipage}[t]{0.38\linewidth}
        \vspace{-2pt}
        \newcolumntype{P}[1]{>{\centering\arraybackslash}p{#1}}
        \newcolumntype{M}[1]{>{\centering\arraybackslash}m{#1}}
        \setlength{\extrarowheight}{4pt}
        \centering
        \resizebox{\linewidth}{!}{%
            \begin{tabular}{@{}M{0.55cm}|M{0.1cm}|M{6.7cm}P{1.85cm}@{}}
                \hline
                \multirow{2}{*}[-4pt]{\rotatebox[origin=c]{90}{\textbf{\renewcommand{\arraystretch}{0.7}\begin{tabular}[c]{@{}c@{}}Aggr-first \\ benefit\end{tabular}}}}                                                                                                                               &
                \rotatetext{FWP}                                                                                                                                                                                                                                                                 &
                \color{blue}$\overbrace{(n_{Src}-n_{Dst})}^{\text{\fontsize{9}{20}\selectfont Reduction factor}}\!\!\!\!\cdot\:$\color{black}$\overbrace{(\color{gray}\alpha\color{black}\cdot{n_{Hid}\cdot{n_{Feat}}}+\color{gray}\beta\color{black}\cdot{n_{Hid}})}^\text{\fontsize{9}{20}\selectfont Kernel execution}$                          &
                \begin{tabular}[c]{@{}c@{}}
                    \color{gray}$\mathbf{\alpha}$\color{black} \textbf{=} $\mathbf{6 \cdot 10^{-5}}$ \\ [-2pt]
                    \color{gray}$\mathbf{\beta}$\color{black} \textbf{=} $\mathbf{1 \cdot 10^{-5}}$
                \end{tabular}                                                                                                                                                                                   \\ \cline{2-4} &
                \rotatetext{BWP}                                                                                                                                                                                                                                                                 &
                \color{blue} $\;\;\:(n_{Src}-n_{Dst})\:\cdot\:$\color{black}$(\color{gray}\alpha\color{black}\cdot{n_{Hid}\cdot{n_{Feat}}}+\color{gray}\beta\color{black}\cdot{n_{Feat}})$                                                                                                                                           &
                \begin{tabular}[c]{@{}c@{}}
                    \color{gray}$\mathbf{\alpha}$\color{black} \textbf{=} $\mathbf{1 \cdot 10^{-7}}$ \\ [-2pt]
                    \color{gray}$\mathbf{\beta}$\color{black} \textbf{=} $\mathbf{4 \cdot 10^{-6}}$
                \end{tabular}                                                                                                                                                                                   \\ \hline
                \multirow{2}{*}[-1pt]{\rotatebox[origin=c]{90}{\textbf{\renewcommand{\arraystretch}{0.7}\begin{tabular}[c]{@{}c@{}}Comb-first \\ benefit \vspace{2pt} \end{tabular}}}}                                                                                                                 &
                \rotatetext{FWP}                                                                                                                                                                                                                                                                 &
                \color{blue}$\overbrace{(n_{Feat}-n_{Hid})}^\text{\fontsize{9}{20}\selectfont Reduction factor}\!\!\cdot\:$\color{black}$\overbrace{(\color{gray}\gamma\color{black} \cdot n_{Edge} + \color{gray}\delta\color{black}\cdot n_{Dst})}^\text{\fontsize{9}{20}\selectfont Kernel execution}$ &
                \begin{tabular}[c]{@{}c@{}}
                    \color{gray}$\mathbf{\gamma}$\color{black} \textbf{=} $\mathbf{1 \cdot 10^{-3}}$ \\ [-2pt]
                    \color{gray}$\mathbf{\delta}$\color{black} \textbf{=} $\mathbf{1 \cdot 10^{-12}}$
                \end{tabular}                                                                                                                                                                                 \\ \cline{2-4} &
                \rotatetext{BWP}                                                                                                                                                                                                                                                                 &
                \color{blue}$\,(n_{Feat}-n_{Hid})\:\cdot\:$\color{black} $(\color{gray}\gamma\color{black} \cdot n_{Edge} + \color{gray}\delta\color{black} \cdot n_{Src})$                                                                                                                                &
                \begin{tabular}[c]{@{}c@{}}
                    \color{gray}$\mathbf{\gamma}$\color{black} \textbf{=} $\mathbf{1\cdot 10^{-6}}$ \\ [-2pt]
                    \color{gray}$\mathbf{\delta}$\color{black} \textbf{=} $\mathbf{1 \cdot 10^{-8}}$
                \end{tabular}                                                                                                                                                                                \\ \hline
            \end{tabular}}
        \vspace{-4pt}
        \captionof{table}{DKP cost model.}
        \label{tbl:cost_model}
        \vspace{-5pt}
    \end{minipage}
    \addtocounter{figure}{-1}
\end{figure*}

\subsection{Frontend for Pure Vertex-centric GNN}
\label{subsec:frontend}
\noindent \textbf{GNN graph aware scheduling.}
The edge-wise scheduling of Graph-approach can be optimal for the conventional graph processing, which exhibits many per-node edges to visit, thereby processing all the edges in parallel.
However, it cannot take advantage of the massive computing power of GPUs when processing the preprocessed graphs for GNNs.
Figure \ref{fig:degree_stat} compares the average \emph{degree} (edge per vertex) of original and preprocessed graphs as well as the corresponding standard deviations.
To examine the detailed degree distributions of both types of graphs, we also analyze the CDF of the degree, as shown in Figures \ref{fig:cdf_original} and \ref{fig:cdf_sampled}.
In these figures, the original graphs exhibit many edges per vertex, but the average degree of the preprocessed graphs is 3.4 times smaller than that of the original graphs.
In addition, the degree of preprocessed graphs is very even.
Thus, in contrast to the existing frameworks that simultaneously process embeddings per edge, it is better for GNNs to parallelize the process of embeddings per node.

To enable high-performance, pure vertex-centric GNN processing, we also need to consider feature vector dimension.
While traditional graph processing usually has a scalar value as a node feature \cite{harish2007accelerating}, GNN models need to process features with a much higher dimension.
Considering such characteristics of features, we traverse graphs and schedule SM threads in a destination-centric, feature-wise manner.
Specifically, as shown in Figure \ref{fig:pure_vertex_thread_sched}, GraphTensor groups all the features associated with each dst node and allocates them to be processed in parallel within the same SM.
This \emph{feature-wise thread scheduling} can maximize the parallelism without memory/cache bloats by being aware of the GNN graph characteristics (i.e., a bounded number of neighbors per dst node and high dimensionality of features).
Also, it does not require COO graph structures and corresponding data processing algorithms, which can realize our pure vertex-centric GNN computing.

\begin{figure}
    \addtocounter{figure}{-1}
    \begin{minipage}{\linewidth}
        \lstset{style=pythonStyle}
        \begin{lstlisting}[]
multiNGCF(nLayers,graph,embed):
  mode$_f$,mode$_g$,mode$_h$ = "mean", "element-wise product", "sum"
  for $\ell\,$ in range(1, nLayers+1):
    CSR$_\ell$ = graph[${\ell\,-1}$]
    @edge@=~NeighborApply~(CSR$_\ell$,embed,mode$_f$)
    @aggr@=~Pull~(CSR$_\ell$,embed,@edge@,mode$_g$,mode$_h$)
    @embed@=~Apply~(@aggr@)
  <return> @embed@
        \end{lstlisting}
    \end{minipage}
    \vspace{-10pt}
    \caption{NAPA programming model.}
    \vspace{-21pt}
    \label{alg:multilayer_ngcf_code}
    \addtocounter{figure}{1}
\end{figure}

\noindent \textbf{Programming model/interfaces.}
Our NAPA programming model mainly consists of three primitives, i) \texttt{NeighborApply}, ii) \texttt{Pull}, and iii) \texttt{Apply}.
\texttt{NeighborApply} and \texttt{Pull} process feature vectors and multiple per-layer subgraphs, represented only in CSR, by completely realizing SDDMM (edge weighting) and SpMM (aggregation).
Compared to DL-approach, NAPA has no sparse-to-dense data conversion since our \texttt{NeighborApply} directly applies $g$ to the embeddings by referring the each layer's subgraph.
In contrast to Graph-approach, our \texttt{NeighborApply} locates all the dst-related embeddings into the same SM and processes them by allocating SM threads in a feature-wise manner.
For example, as NAPA traverses the target graph based on dst nodes (destination-centric),
\texttt{NeighborApply} logically splits all the currently visiting nodes (e.g., $V_0$, $V_2$, and $V_3$) into multiple sub-embeddings and allocates each of them to different SMs (Figure \ref{fig:pure_vertex_neighbor_apply}).
Note that Graph-approach's edge weighting redundantly copies the dst node's embedding as many as its src nodes across multiple SMs.
In contrast, NAPA loads dst nodes' embedding only once and reuses the embedding during \texttt{NeighborApply}.
On the other hand, \texttt{Pull} loads the weights computed by \texttt{NeighborApply} and the corresponding src node's embeddings (Figure \ref{fig:pure_vertex_pull}).
It then applies $f$ to the embeddings in the feature-wise manner.
Thus, when \texttt{Pull} accumulates the embeddings of $V_2$ and $V_3$, the target SM can recycle the output of $h$ (weighted embedding). % such that it removes the memory accesses for the aggregation.
In addition, similar to \texttt{NeighborApply}, \texttt{Pull} reuses the output embeddings when $f$ accumulates all the target embeddings.
By reusing the embeddings in SMs, NAPA can reduce the global memory accesses, thereby shortening the execution time further.
For the combination, MLP computations are mostly dense matrix transformation, which is already well harmonized with GPU's massive computing.
Thus, our \texttt{Apply} leverages the primitives of TensorFlow (e.g., \texttt{tf.matmul}, \texttt{tf.nn.relu}, and \texttt{tf.nn.bias\_add}) to implement conventional MLP.

Using our NAPA programming model, users can implement diverse GNN models.
Algorithm \ref{alg:multilayer_ngcf_code} shows an example of the high-level implementation of \emph{neural graph collaborative filtering} (NGCF \cite{wang2019ngcf}) that computes edge weights in addition to a basic aggregation and combination.
The users first configure the type of functions, $f$, $g$, and $h$, such as mean, element-wise product, and sum (for an average-based aggregation), using a \texttt{mode} variable (lines 2-3).
For a given number of layers (\texttt{nLayers}), the NGCF model iteratively calculates edge weights (\texttt{Edge}), accumulates them with the target embeddings (\texttt{embed}), and transforms the aggregated results (\texttt{aggr}) to the embedding result (lines 6-8).
In the meantime, NGCF's each layer can retrieve the corresponding subgraph from the input graph (\texttt{graph}) by using its layer index, such that \texttt{NeighborApply}, \texttt{Pull}, and \texttt{Apply} can process the appropriate subgraph data and embeddings.
Note that users can simply apply different GNN models by reconfiguring the \texttt{modes}.

%-------------------------------------------------------------------------------
\section{Backend Kernel and Tensor Management}
%-------------------------------------------------------------------------------
\label{sec:implementation}
\begin{figure*}[t]
  \begin{minipage}[t]{0.34\linewidth}
      \includegraphics[width=0.98\linewidth, valign=t]{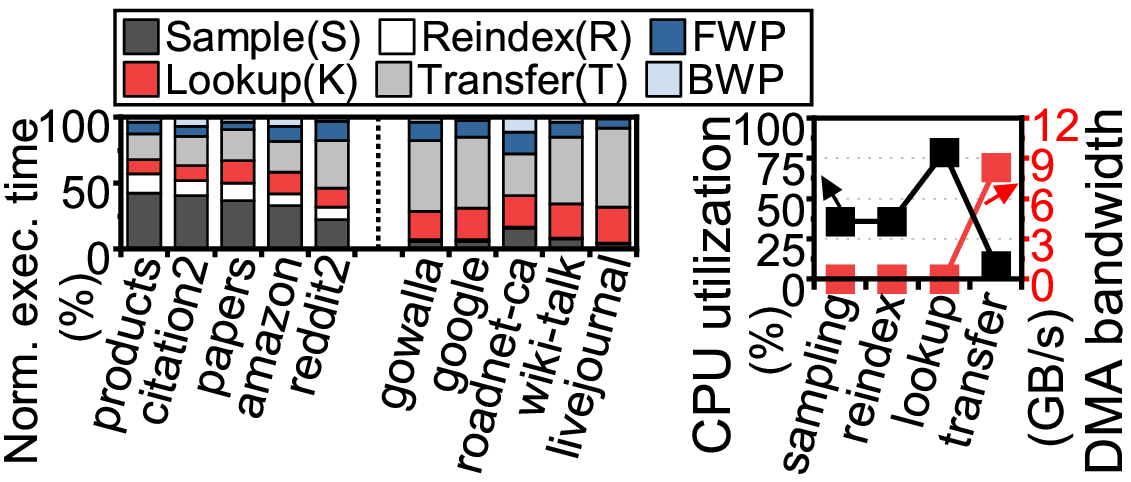}

      \vspace{2pt}
    \hspace{0.05\linewidth}
    \begin{subfigure}[t]{.3\linewidth}
      \vspace{-8pt}
      \caption{End-to-end\\breakdown.}
      \label{fig:chal_e2e_breakdown}
    \end{subfigure}
    \hspace{0.05\linewidth}
    \begin{subfigure}[t]{.55\linewidth}
      \vspace{-8pt}
      \caption{System resource\\utilization (\texttt{wiki-talk}).}
      \label{fig:chal_system_util}
    \end{subfigure}
    \vspace{-7pt}
    \caption{End-to-end performance.}
    \label{fig:csr2csc}
  \end{minipage}
  \begin{minipage}[t]{0.65\linewidth}
    \begin{minipage}[t]{0.43\linewidth}
      \vspace{-13pt}
      \includegraphics[width=1\linewidth, valign=t]{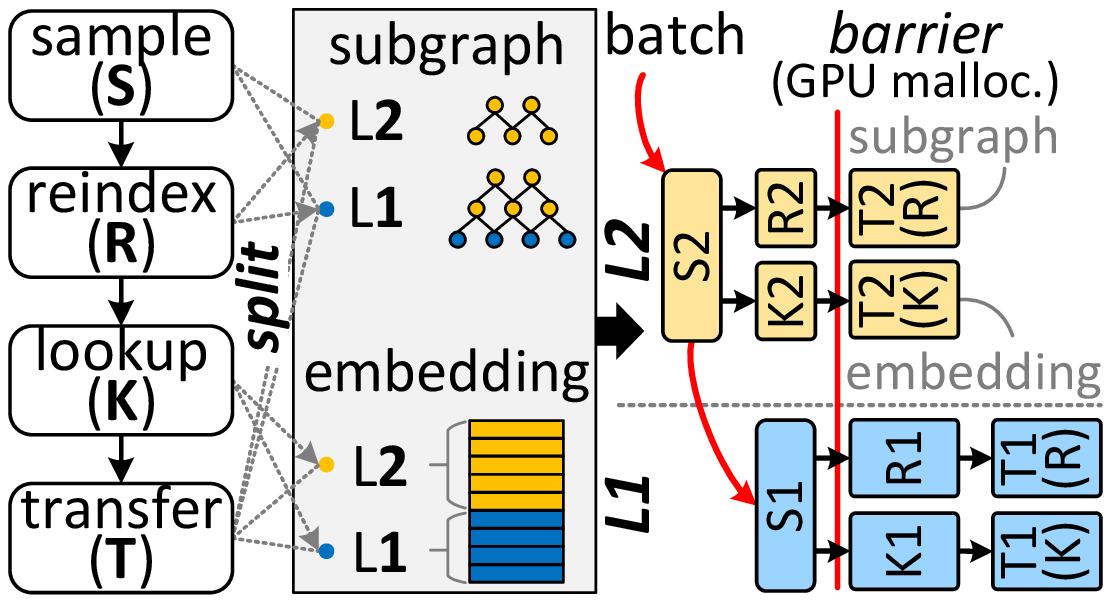}
    \end{minipage}
    \begin{minipage}[t]{0.55\linewidth}
      \vspace{-13pt}
      \includegraphics[width=0.26\linewidth, valign=t]{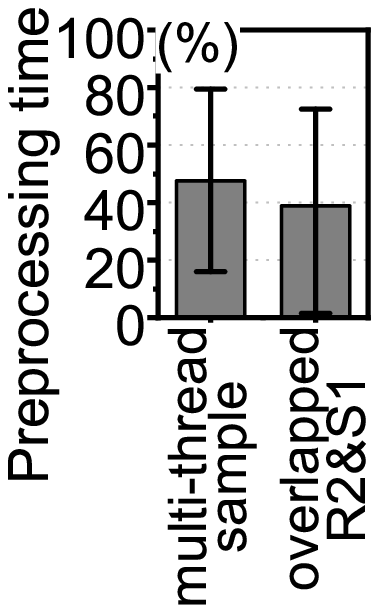}
      \includegraphics[width=0.72\linewidth, valign=t]{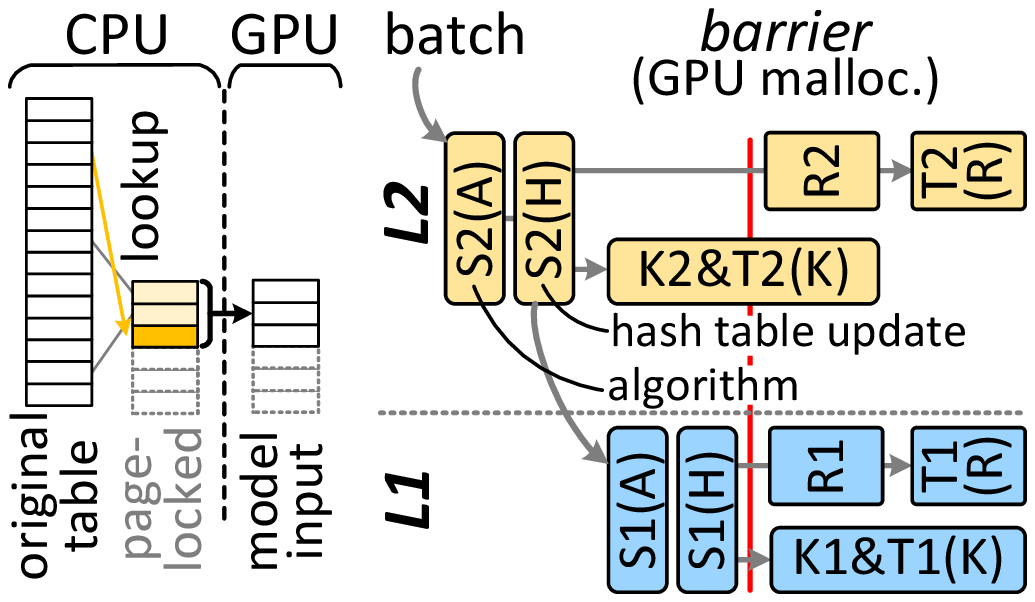}
    \end{minipage}

    \hspace{0.01\linewidth}
    \begin{minipage}[t]{0.34\linewidth}
      \begin{subfigure}{\linewidth} % Blank subfigure to resolve reference error
      \end{subfigure}
      \vspace{-5pt}
      \caption{Service-wide tensor scheduling.}
      \label{fig:design_scheduling}
    \end{minipage}
    \hspace{0.01\linewidth}
    \begin{minipage}[t]{0.6\linewidth}
      \begin{subfigure}[t]{0.27\linewidth}
        \vspace{-4pt}
        \caption{Contention.}
        \label{fig:chal_lock}
      \end{subfigure}
      \begin{subfigure}[t]{0.22\linewidth}
        \vspace{-4pt}
        \caption{Overlap.}
        \label{fig:design_overlap}
      \end{subfigure}
      \begin{subfigure}[t]{0.48\linewidth}
        \vspace{-4pt}
        \caption{Optimized scheduling.}
        \label{fig:design_pre2}
      \end{subfigure}
      \vspace{-5pt}
      \caption{Contention relaxing.}
      \label{design_pre}
    \end{minipage}
  \end{minipage}
  \vspace{-20pt}
\end{figure*}

\subsection{GNN Kernel Orchestrator}
The kernel orchestrator's goal is to reduce the embedding dimensionality such that the GNN latency can be shortened.

\noindent \textbf{Dimensionality reduction.}
Figure \ref{fig:dkp_concept} explains how the dimensions of embeddings are changed by aggregation and combination.
Since the aggregation accumulates the $n_{Src}$ number of src's embeddings to the $n_{Dst}$ number of dst's embeddings, the dimension of aggregated embeddings becomes equal to $n_{Dst}$. % without a change of the $n_{Feature}$ number of features.
In contrast, the combination transforms the $n_{Feature}$ number of features to the $n_{Hidden}$ columns of the hidden layer's weight matrix of MLP, reducing $n_{Feature}$ rather than $n_{Src}$.
Because of these characteristics, the size of the input tensor for FWP ($f$ and $MLP$) and BWP ($f'$ and $MLP'$) can vary based on which function is applied first.
To be precise, we analyze the dimensionality reduction rate (per GNN layer) for the workloads that we tested in \cref{sec:evaluation} (Figure \ref{fig:dkp_motiv}).
When we run the combination first (rather than the aggregation), all layers of \texttt{wiki-talk} can reduce the input tensor size by 31.7\%, on average,
while other layers can still take advantage by following the conventional execution order.

\noindent \textbf{Dynamic kernel placement (DKP).}
Our kernel orchestrator's DKP rearranges the aggregation and combination kernels if the combination can reduce the total amount of features more than the aggregation.
To this end, as shown in Figure \ref{fig:dkp_impl}, our kernel orchestrator checks the \emph{dataflow graph} (DFG) at runtime, searching for NAPA's \texttt{Pull} and the subsequent matrix multiplication of MLP (\texttt{MatMul}).
Since it is prohibited to change the execution sequence of delegated kernels at the GPU-side, the kernel orchestrator prepares a new DFG node (\emph{Cost-DKP}) in advance, and replaces the two nodes with it at the host-side.
Then, it disconnects the links associated with \texttt{Pull}'s input and MatMul's output from the original nodes and links them to our Cost-DKP (and MLP's bias).
At runtime, Cost-DKP examines the input tensor's dimensionality and performs the combination first if its reduction rate is higher than the original execution sequence.
We model the kernel latency based on the numbers of nodes and embeddings, and introduce such a cost model into the Cost-DKP for its kernel rearrangement at runtime, which will be explained shortly.
Note that, as discussed in \textsection \ref{subsec:graph_neural_networks}, the \texttt{Pull} and \texttt{Matmul} that we rearrange are commonly observed in all the existing GNN models.
This characteristic makes our DKP robust and applicable to diverse GNN models.

We also show how the combination can be executed first at the bottom of Figure \ref{fig:dkp_impl}.
The aggregation-first FWP processing can be simplified by $MLP(f(h(X)))=\sigma(Wf(h(X))+b)$ where $\sigma$, $W$, and $b$ are a non-linear function, MLP's weight, and bias, respectively.
As the aggregation $f(h(X))$ can be rewritten as $XA^T$ (SpMM), where $A$ is the sparse matrix, we can get an equation of $\sigma((WX)A^T+b)=\sigma(f(h(WX))+b)$.
From the equation, combination can be scheduled first by calculating MLP's MatMul, earlier than $h$ and $f$.
Similarly, we can execute BWP with different execution order based on chain rules and matrix algebra's associative property.

\noindent \textbf{DKP cost model and parameters.}
While it is clear to see the benefits of changing the execution order of aggregation and combination kernels from Figure \ref{fig:dkp_motiv},
it is non-trivial to determine the best order before their execution.
Thus, we introduce a cost model that estimates the kernel latency based on the dimensionality of input tensors.
Table \ref{tbl:cost_model} explains the cost model that estimates the benefits (in terms of latency) for the aggregation-first and combination-first kernel placements.
The model consists of i) reduction factor and ii) kernel execution factor.
Since the aggregation reduces the input height of the following combination from $n_{Src}$ to $n_{Dst}$, we estimate its cost as $n_{Src}-n_{Dst}$; the kernel time is proportional to the reduced input and $n_{Hidden}$.
On the other hand, the combination reduces the input width of the aggregation from $n_{Feature}$ to $n_{Hidden}$; its reduction factor can be $n_{Feature}-n_{Hidden}$.
In this case, the kernel execution factor considers the memory access time for dst nodes. % (owing to our destination-centric graph traversing and feature-wse kernel scheduling)
Thus, the latency of the combination-first kernel placement can be proportional to the dimensionality of transformed embeddings, $n_{Edges}$, and $n_{Dst}$.
The cost model for BWP is largely similar to that of FWP, while some terms are replaced considering their different directions of graph traversing and matrix transpose (e.g., $n_{Hid} \rightarrow n_{Feat}$ and $n_{Dst} \rightarrow n_{Src}$).
However, when we estimate the aggregation-first's benefit of BWP for the first GNN layer (which is the last in terms of execution order), we set its reduction factor as $n_{Src}$.
This is because the aggregation-first's BWP does not need to perform aggregation's BWP for calculating the gradient for MLP parameters, thereby having a bigger benefit.
Note that the only goal of training is to update MLP parameters by calculating the gradient of the parameters.

Since the cost model's coefficient parameters can vary based on system performance, DKP fits the parameters by leveraging least-squares estimation algorithm \cite{van2005least} with the measured kernel execution time.
This coefficient fitting happens at the beginning of the first epoch in training and the result coefficients are leveraged until the training ends.
Note that since GNNs iterate the same computing procedure with different model parameters and sampled batches, the target model execution latency across different epochs is sustained.
The results are given at the right of the table.
In our preliminary evaluations using those parameters, the estimated times are close to the actual latency (only 12.5\% error). %under the processing of diverse workloads that we tested in \cref{sec:evaluation}.

\subsection{Service-wide Tensor Scheduling}
\label{subsec:service-wide-tensor-sched}
Even though we can shorten the latency of GNN computing, users can experience low performance because of preprocessing latency.
Note that GPUs can start training only if the input subgraphs/embeddings are ready in the device.
For better understanding, we decompose end-to-end latency into each preprocessing component and GNN computing time.
As shown in Figure \ref{fig:chal_e2e_breakdown}, the latency of GNN computing (FWP+BWP) only accounts for 15.8\% of the end-to-end latency, on average.
In general, the neighbor sampling ($S$) task dominates the preprocessing time for the workloads with a relatively small size of features (i.e., the left half of the figure).
In contrast, the data preparation, including reindexing ($R$), embedding lookup ($K$), and data transfer ($T$) tasks consumes most of the preprocessing latency for the workloads with heavy features (i.e., the right half of the figure).

Unfortunately, it is non-trivial to make individual preprocessing tasks' latency shorter since they mainly consist of random memory access or data transfer through PCIe.
On the other hand, existing frameworks waste useful system resources due to their serialized execution.
As shown in Figure \ref{fig:chal_system_util}, $S$, $R$, and $K$ tasks do not require data transfer between host and GPU, thereby not utilizing PCIe resources during their execution.
In contrast, the $T$ task uses only a single core while leaving other multiple cores idle.

\noindent \textbf{High-performance preprocessing.}
Instead of shortening the individual latency of preprocessing tasks, our service-wide tensor scheduler parallelizes and pipelines the tasks to maximize the resource utilization.
It splits each component into multiple subtasks while being aware of data type to transfer and dependency of each subtask.
The executions of the subtasks are then parallelized across multiple threads (per node). % by allocating the threads per the target dst node.
In cases where we cannot parallelize a subtask, the scheduler makes several subtasks without dependency run together.

\begin{figure*}[b!]
  \vspace{-14pt}
    \includegraphics[width=1\linewidth]{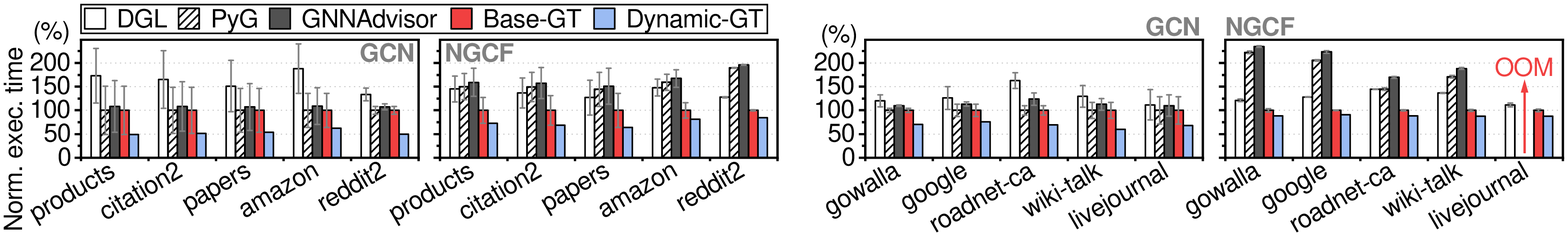}

    \vspace{-10pt}
    \begin{subfigure}{.49\linewidth}
      \vspace{-14pt}
      \caption{Light feature graphs.}
      \label{fig:eval_inference_light}
      \vspace{-7pt}
    \end{subfigure}
    \begin{subfigure}{.5\linewidth}
      \vspace{-14pt}
      \caption{Heavy feature graphs.}
      \label{fig:eval_inference_heavy}
      \vspace{-7pt}
    \end{subfigure}

    \vspace{-2pt}
    \caption{Training latency analysis.}
    \vspace{-10pt}
    \label{fig:eval_inference}
\end{figure*}

\begin{figure}
  \centering
  \small
  \setlength{\tabcolsep}{2pt}
  \renewcommand{\arraystretch}{.97}
    \resizebox{1\linewidth}{!}{%
    \begin{threeparttable}
      \begin{tabular}{@{}|cp{4.5em}|r|r|r|r|r|r|r|r|r|@{}}
        \hline
            &   & \multicolumn{3}{c|}{\textbf{Full graph}} & \multicolumn{5}{c|}{\textbf{Sampled graph}} & \multicolumn{1}{c|}{\textbf{Task}} \\ \cline{3-11}
            & \multicolumn{1}{c|}{Name}
                & \multicolumn{1}{c|}{\begin{tabular}[c]{@{}c@{}}Ver-    \\[-1pt] tices \end{tabular}}
                & \multicolumn{1}{c|}{Edges}
                & \multicolumn{1}{c|}{\begin{tabular}[c]{@{}c@{}}Feature\\[-1pt] dim.\end{tabular}}
                & \multicolumn{1}{c|}{\begin{tabular}[c]{@{}c@{}}Ver-    \\[-1pt] tices \end{tabular}}
                & \multicolumn{1}{c|}{Edges}
                & \multicolumn{1}{c|}{\begin{tabular}[c]{@{}c@{}}Dst.    \\[-1pt] vertices \end{tabular}}
                & \multicolumn{1}{c|}{\begin{tabular}[c]{@{}c@{}}Edges / \\[-1pt] vertices \end{tabular}}
                & \multicolumn{1}{c|}{\begin{tabular}[c]{@{}c@{}}Feature   \\[-1pt] size \end{tabular}}
                & \multicolumn{1}{c|}{\begin{tabular}[c]{@{}c@{}}Out.    \\[-1pt] dim.\end{tabular}} \\ \hline
                %& \multicolumn{1}{c|}{\begin{tabular}[c]{@{}c@{}}Base-    \\[-1pt] GT\end{tabular}} \\ \hline
        \multirow{5}{*}{\rotatebox[origin=c]{90}{\textbf{Light feat.}}}
                & products\tnote{1}    &   2M & 124M &   100 &  351K &  767K &  50K &  2.2 & 140\scriptsize{MB} & 47 \\ [-1pt]
                & citation2\tnote{1}   &   3M &  61M &   128 &  322K &  592K &  41K &  1.8 &  165\scriptsize{MB} & 2 \\ [-1pt]
                & papers\tnote{1}      & 111M &   2G &   128 &  564K &  751K &  50K &  1.3 &  289\scriptsize{MB} &  172 \\ [-1pt]
                & amazon\tnote{2}      &   2M & 264M &   200 &  154K &  425K &  28K &  2.8 &  124\scriptsize{MB} &  2 \\ [-1pt]
                & reddit2\tnote{2}     & 233K &  23M &   602 &  185K &  912K &  57K &  4.9 &  446\scriptsize{MB} & 41 \\ [-1pt] \hline
        \multirow{5}{*}{\rotatebox[origin=c]{90}{\textbf{Heavy feat.}}}
                & gowalla\tnote{3}     & 197K &   2M &  4353 &  54K &  183K & 15K &  3.4 & 943\scriptsize{MB} &  2 \\ [-1pt]
                & google\tnote{3}      & 916K &   5M &  4353 &  54K &  177K & 16K &  3.3 & 936\scriptsize{MB} &  2 \\ [-1pt]
                & roadnet-ca\tnote{3}  &  2M  &   6M &  4353 &   5K &   17K &  4K &  3.3 &  23\scriptsize{MB} &  2 \\ [-1pt]
                & wiki-talk\tnote{3}   &  2M  &   5M &  4353 &  29K &   60K &  8K &  2.1 & 497\scriptsize{MB} &  2 \\ [-1pt]
                & livejournal\tnote{3} &  5M  &  96M &  4353 & 233K &  393K & 28K &  1.7 & 4\scriptsize{GB} &  2 \\ [-1pt] \hline
      \end{tabular}
      \begin{tablenotes}
        \item[1] OGB \cite{hu2020ogb}\hspace{-14pt} \item[2] GraphSAINT \cite{graphsaint-iclr20} \hspace{-14pt} \item[3] SNAP \cite{snapnets} \hspace{-10pt}
        \item[*] dim.=dimension, out.=output
      \end{tablenotes}
    \end{threeparttable}
  }
  \vspace{-5pt}
  \captionof{table}{Important characteristics of graphs.}
  \vspace{-18pt}
  \label{tbl:dataset}
\end{figure}

Figure \ref{fig:design_scheduling} shows the basic idea of our service-wide tensor scheduling; in this example, we assume that the GNN model employs two layers. %for the sake of brevity.
The scheduler subdivides a single, large preprocessing task into $S$, $R$, $K$, and $T$ per-layer subtasks, and it further classifies the subtasks based on its input data type (e.g., subgraphs, embeddings).
Since the data preparation ($R$, $K$, $T$) can be performed \emph{only if} the sampled graphs are ready, we schedule \texttt{S2} and \texttt{S1} subtasks (for layers 2 and 1, respectively) back-to-back.
Note that the data preparation renumbers the subgraph and retrieves the embeddings for the sampled subgraph.
Further, we schedule $R$, $K$ subtasks for the layer 2 before the completion of \texttt{S1}, such that we can execute them with the maximum parallelism.
During this time, the scheduler runs $R$ and $K$ together as they have no dependency due to their different input data type.
The $T(R)$ and $T(K)$ are also run in parallel, due to the same reason.
Note that the size of subgraphs and embedding tables that $T$ need for their memory allocation is determined after \texttt{S1}.
Thus, the scheduler sets a barrier before running $T$ that waits for $S1$'s completion.

\noindent \textbf{Relaxing contention.}
While the tensor scheduling can reduce the preprocessing times, its thread-level parallelism is limited because of lock contention.
Since S and R tasks require updating/referring the hash table for all the sampled nodes, which is a shared resource, they are in a race condition for the hash table accesses.
As shown in Figure \ref{fig:chal_lock}, the contention overhead caused by the threads in $S$ subtasks (\texttt{S1}/\texttt{S2}) and between $S$ and $R$ subtasks accounts for 47.4\% and 39.0\% of the total preprocessing times.
To address this lock contention, we further divide $S$ subtasks into two parts, each being involved in the algorithm execution ($A$) and hash table updates ($H$) (Figure \ref{fig:design_pre2}).
We then serialize $S$ subtasks dealing with the hash table updates while fully parallelizing the algorithm part.
Similarly, we avoid overlapping several subtasks competing to access the shared resources (e.g., $S$ and $R$).
Lastly, to further overlap the time for $K$ and $T$ subtasks, the service-wide tensor scheduler immediately transfers each sampled embedding whenever it is ready on a buffer (by $K$ subtasks) in a pipelined manner (Figure \ref{fig:design_overlap}).
Since we always transfer the embeddings in K's output buffer, we allocate the buffer on page-locked memory (pinned memory in CUDA) to avoid unnecessary data copy by GPU driver.
We also overlap the execution time of preprocessing with FWP/BWP in GPU, which is a common practice for the existing deep learning frameworks such as Tensorflow and Pytorch.

%-------------------------------------------------------------------------------
\section{Evaluation}
%-------------------------------------------------------------------------------
\label{sec:evaluation}
\noindent \textbf{Evaluation method.}
For the evaluation of DL-approach and Graph-approach, we use PyG 1.7.0 \cite{iclr19pyg} and DGL 0.8.2 \cite{wang2019dgl}.
We build three different versions of GraphTensor; %, i) \emph{Base-GT}, ii) \emph{Dynamic-GT}, and iii) \emph{Prepro-GT}.
i) \emph{Base-GT} is the baseline that employs NAPA, but has no DKP, whereas ii) \emph{Dynamic-GT} and iii) \emph{Prepro-GT} employ DKP and DKP with service-wide tensor scheduler, respectively.
All these GraphTensor leverage TensorFlow 2.4.0 with CUDA 11.1 and CUDNN 8.
We use NVIDIA RTX 3090 GPU, which has 82 1.4GHz SM processors with 24GB GDDR6X DRAM.
The testbed employs a 3.0GHz 12 core processor (Intel Xeon Gold 5317) and 196GB DDR4-2933 main memory.

\begin{figure*}[t]
  \centering
  \vspace{-5pt}
  \begin{minipage}[t]{0.43\linewidth}
    \vspace{0pt}
    \includegraphics[width=1\linewidth]{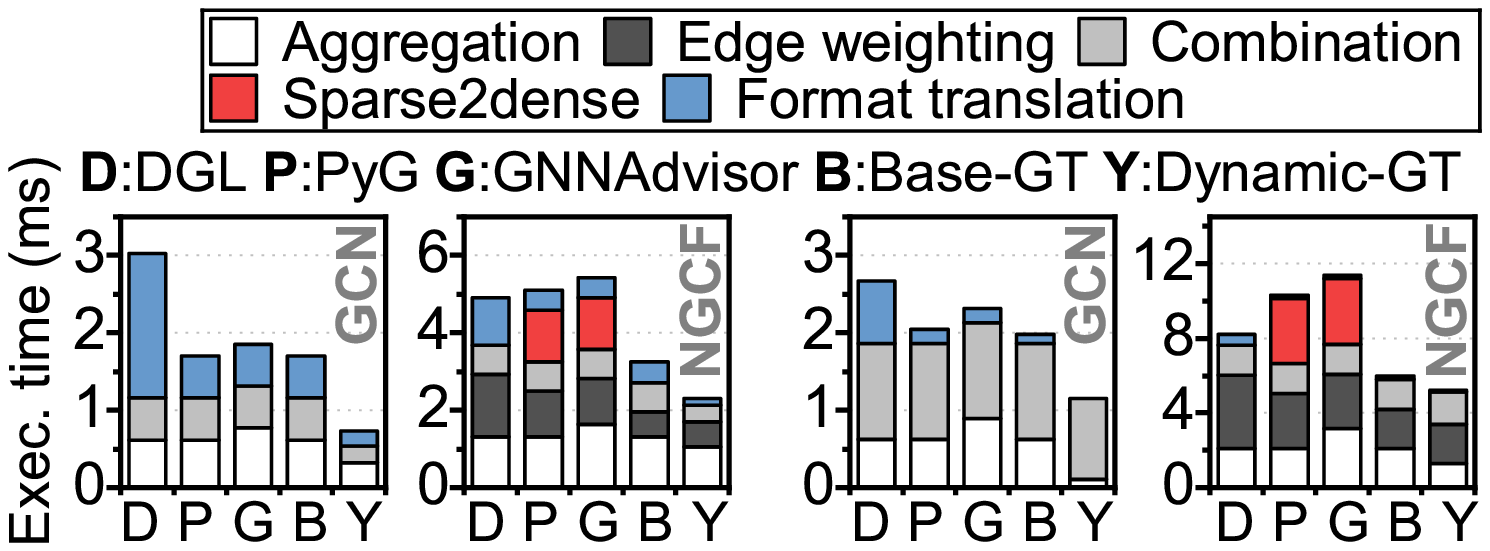}
    \vspace{-8pt}
    \hspace{0.01\linewidth}
    \begin{subfigure}{.46\linewidth}
      \vspace{-14pt}
      \caption{Light (\texttt{products}).}
      \label{fig:eval_breakdown_light}
      \vspace{2pt}
    \end{subfigure}
    \begin{subfigure}{.5\linewidth}
      \vspace{-14pt}
      \caption{Heavy (\texttt{wiki-talk}).}
      \label{fig:eval_breakdown_heavy}
      \vspace{2pt}
    \end{subfigure}

    \caption{GPU kernel execution breakdown.}
    \label{fig:eval_inference_breakdown}
    \vspace{-7pt}
  \end{minipage}
  \begin{minipage}[t]{0.345\linewidth}
    \strut\vspace*{-\baselineskip}\newline

    \includegraphics[width=1\linewidth]{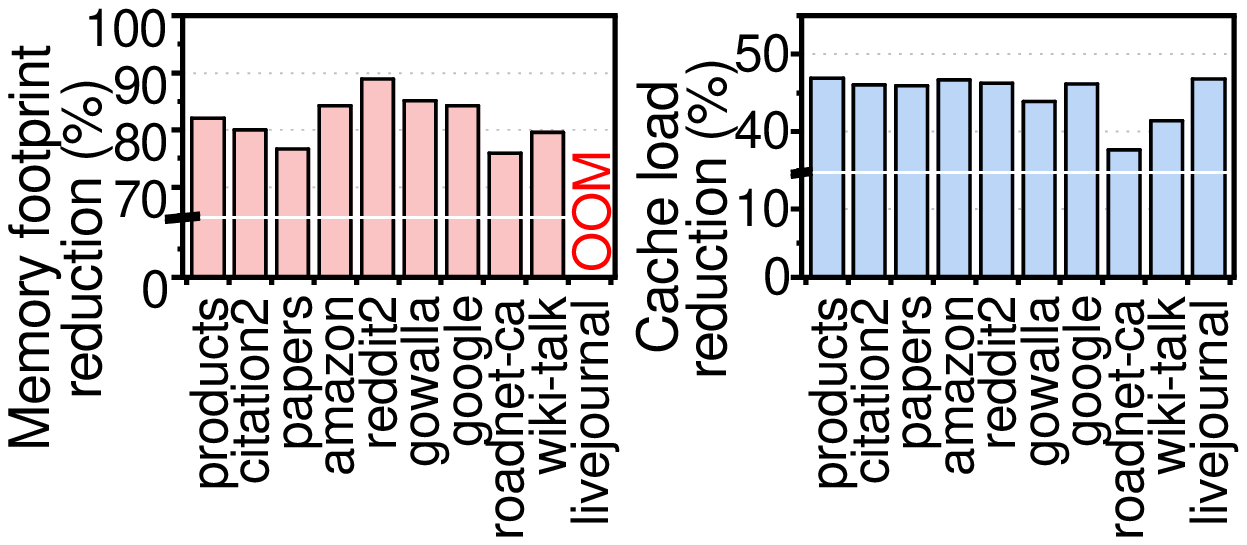}

    \vspace{-1.5pt}
    \hspace{0.01\linewidth}
    \begin{subfigure}{.46\linewidth}
      \vspace{-12pt}
      \caption{Memory usage.}
      \label{fig:eval_membloat}
      \vspace{-2pt}
    \end{subfigure}
    \begin{subfigure}{.5\linewidth}
      \vspace{-12pt}
      \caption{Cache usage.}
      \label{fig:eval_cachebloat}
      \vspace{-2pt}
    \end{subfigure}

    \vspace{-3.6pt}
    \caption{NAPA's GPU resource usage.}
    \label{fig:eval_gpu_resource}
    \vspace{-18pt}
  \end{minipage}
  \begin{minipage}[t]{0.205\linewidth}
    \strut\vspace*{-\baselineskip}\newline

    \includegraphics[width=1\linewidth]{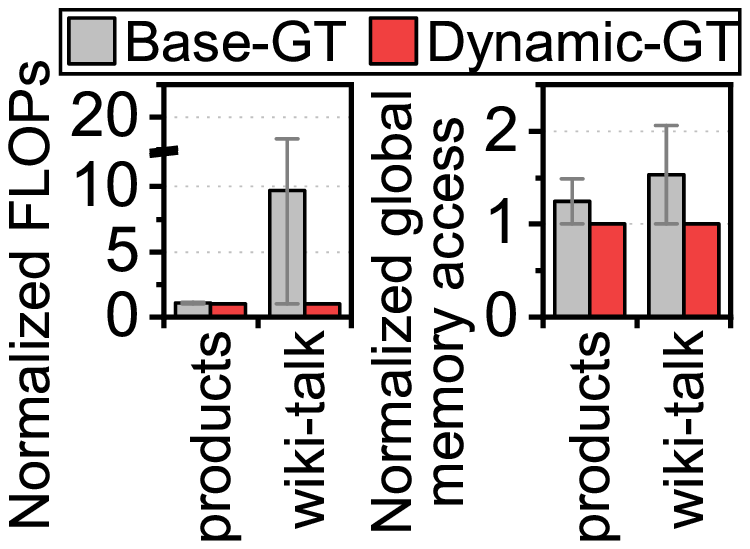}

    \vspace{-5pt}
    \begin{subfigure}{1\linewidth}
      \centering
      \renewcommand*{\arraystretch}{0.3}
      \setlength{\tabcolsep}{3pt}
      \begin{tabularx}{\textwidth}{
        p{\dimexpr.49\linewidth-2\tabcolsep-1.3333\arrayrulewidth}% column 2
        p{\dimexpr.49\linewidth-2\tabcolsep-1.3333\arrayrulewidth}% column 2
        }
           \caption{FLOPs.} \label{fig:eval_dkpflop}
        & \caption{Memory.} \label{fig:eval_dkpmem}
      \end{tabularx}
    \end{subfigure}

    % \vspace{-2.6pt}
    % \begin{subfigure}{.46\linewidth}
    %   \vspace{-15pt}
    %   \caption{FLOPs.}
    %   \label{fig:eval_dkpflop}
    %   \vspace{2pt}
    % \end{subfigure}
    % \begin{subfigure}{.5\linewidth}
    %   % \vspace{-2pt}
    %   \caption{Memory.}
    %   \label{fig:eval_dkpmem}
    %   \vspace{2pt}
    % \end{subfigure}

    \vspace{-16pt}
    \caption{DKP analysis.}
    \label{fig:eval_dkpanalysis}
    \vspace{-25pt}
  \end{minipage}

  \vspace{-10pt}
\end{figure*}

\noindent \textbf{GNN models.}
We evaluate two representative GNN models, \emph{Graph convolutional network} (GCN \cite{kipf2017semi}) and NGCF \cite{wang2019ngcf}.
GCN is broadly adopted in many applications, such as node/graph classifications \cite{yao2019graph, you2018graphmolecule}. %\cite{yao2019graph, marcheggiani2017encoding, you2018graphmolecule, zhao2019t, li2019semi}.
GCN simply accumulates the neighbor nodes using an average-based aggregation (Mean) for $f$, whereas it does not weight any edges. %in the target graph.
On the other hand, NGCF is popularly used in recommendation systems \cite{wu2020graph}. %\cite{wu2020graph, lei2020interactive, he2020lightgcn}.
NGCF considers a similarity among embeddings of neighbors by adding a similarity weight to the average-based aggregation.
The similarity weight is calculated by the element-wise product ($g$) and sum-based weight accumulation ($h$) for the aggregation, thereby highlighting the embedding of the nodes with a high similarity score.
\cite{chen2018fastgcn, icml18jumpingknowledge}
are a variation of GCN, while \cite{velickovic2018graph, song2019session}
are similar to NGCF.
Both GCN and NGCF have hidden dimensions of 64.

\noindent \textbf{Workloads and datasets.}
We use 10 graph datasets \cite{hu2020ogb, graphsaint-iclr20, snapnets} for the evaluation, which are popularly used in machine learning community.
There are several graphs that do not provide the input embedding table.
For these workloads, we create the embeddings whose dimensionality is the same as what the industry uses for their GNN adoption \cite{ying2018graph}.
In all experiments, a batch includes 300 vertices.
The characteristics of the graphs and corresponding sampled graphs are explained in Table \ref{tbl:dataset}.
In the table, we classify the workloads whose feature dimension is lower than 4K as \emph{light feature graphs}; the remaining workloads (the feature dimension > 4K) is called \emph{heavy feature graphs}.
We evaluate the frameworks' GNN training performance on these datasets.
Since training simply iterates to precess batches in a given dataset, we show the performance of processing a single batch as a representative of the entire training performance.

\subsection{Performance Analysis.}

Figures \ref{fig:eval_inference_light} and \ref{fig:eval_inference_heavy} compare GNN training latencies of diverse frameworks we tested for light and heavy feature graphs, respectively.
In this evaluation, we also compare a state-of-the-art GNN framework, GNNAdvisor\footnote{To ensure a fair comparison, we disabled GNNAdvisor's additional preprocessing of the input graph (node renumbering);
This overhead is significant for sampling-based training which generates a new input graph for every batch.}
\cite{osdi21gnnadvisor} with other frameworks.
To better understand, we normalize the batch latency of the frameworks to that of Base-GT;
For a fair comparison, we measure the latency of GPU kernel execution using Nsight Systems \cite{nsight-systems}, which excludes framework-specific overhead such as existing DL framework handling.
To dig deeper, we also decompose the latency of two representative workloads, \texttt{products} (light features) and \texttt{wiki-talk} (heavy features), into aggregation, edge weighting, combination, sparse-to-dense data conversion (Sparse2Dense), and GPU format translation, in Figure \ref{fig:eval_inference_breakdown}.
In addition, while DGL, PyG and GNNAdvisor execute the aggregation first and then execute the combination in default, it is possible to explicitly program the combination-first execution based on user's heuristic on system hyperparameters.
Thus, the figures show the average latency of the two execution, while the error bars indicate their individual latency.

\noindent \textbf{Light feature graphs.}
For the light feature graphs (Figure \ref{fig:eval_inference_light}), GCN (left) and NGCF (right) exhibit different performance behaviors in their training.
The performance of DGL (Graph-approach) is 1.6$\times$ worse than Base-GT.
The reason behind this poor performance is the format translation (COO to CSR) for FWP aggregation.
Consider Figure \ref{fig:eval_breakdown_light} to better understand; the format translation accounts 64.5\% and 24.9\% of the total execution time of DGL's GCN and NGCF, respectively.
Note that, format translation requires allocating additional buffers in GPU memory to sort the given edges and calculate the pointers, which also incurs additional overhead.
DGL somehow alleviates such translation overhead for GCN when the target graphs exhibit long feature lengths (\texttt{reddit2}), as the long feature lengths increase the amount of computation for processing the features while format translation overhead stays similar.
Even though COO is the best for edge-centric scheduling to enhance the performance of edge weighting, it is ill-tuned to accumulate embeddings using SpMM.

\begin{figure*}[t]
  \vspace{-5pt}
  \begin{minipage}[t]{0.6\linewidth}
    \strut\vspace*{-\baselineskip}\newline

    \includegraphics[width=\linewidth, valign=t]{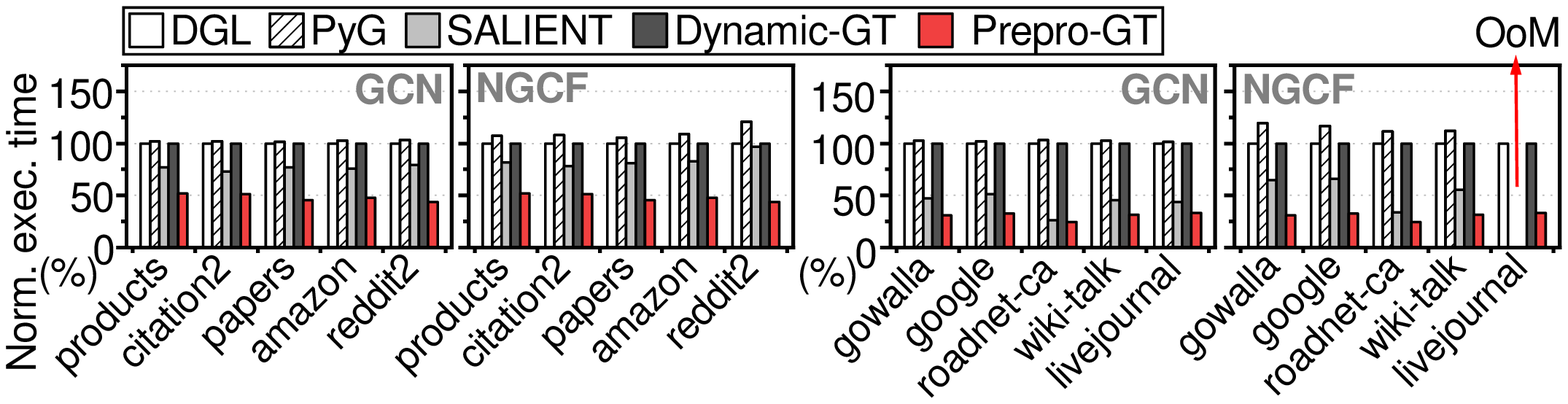}

    \hspace{0.1\linewidth}
    \begin{subfigure}{0.4\linewidth}
      \vspace{-6pt}
      \caption{Light feature graphs.}
      \label{fig:eval_training_light}
    \end{subfigure}
    \hspace{0.03\linewidth}
    \begin{subfigure}{0.4\linewidth}
      \vspace{-6pt}
      \caption{Heavy feature graphs.}
      \label{fig:eval_training_heavy}
    \end{subfigure}
    \vspace{-7pt}
    \caption{End-to-end latency analysis.}
    \label{fig:e2e_exec_time}
  \end{minipage}
  \begin{minipage}[t]{0.39\linewidth}
    \strut\vspace*{-\baselineskip}\newline

    \includegraphics[width=\linewidth, valign=t]{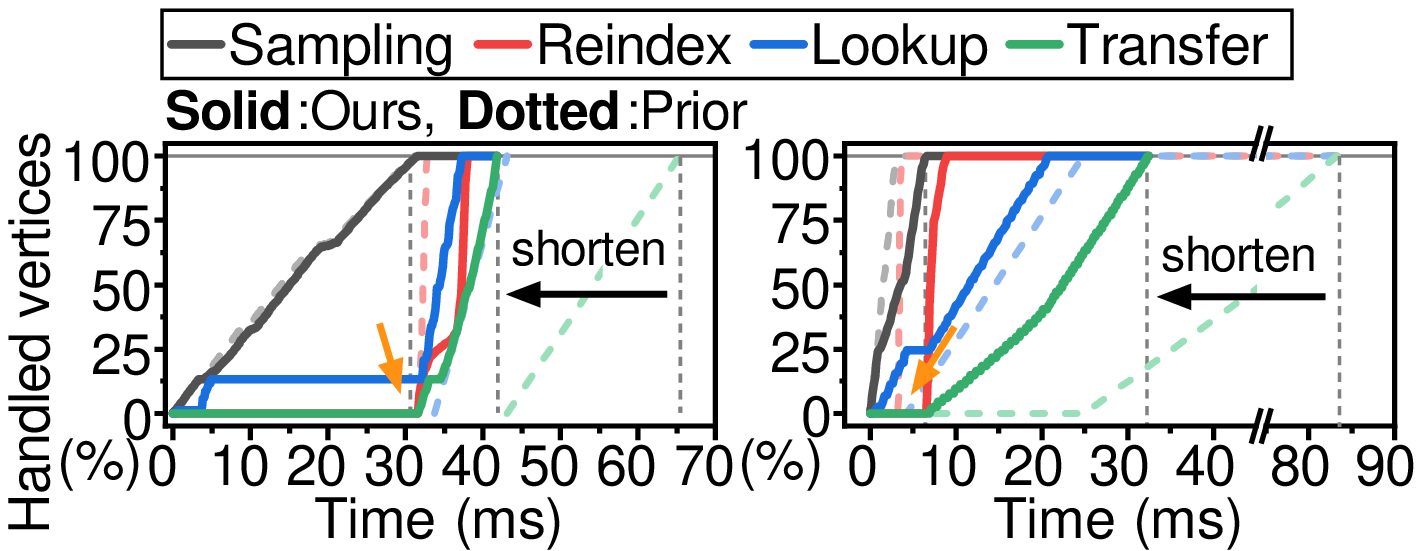}

    \begin{subfigure}{0.5\linewidth}
      \vspace{-15pt}
      \caption{Light (\texttt{products}).}
      \label{fig:eval_timeline_light}
    \end{subfigure}
    \begin{subfigure}{0.47\linewidth}
      \vspace{-6pt}
      \caption{Heavy (\texttt{wiki-talk}).}
      \label{fig:eval_timeline_heavy}
    \end{subfigure}
    \vspace{-7pt}
    \caption{Preprocessing timeline analysis.}
    \label{fig:eval_prepro_timeline}
  \end{minipage}
  \vspace{-20pt}
\end{figure*}

On the other hand, PyG (DL-approach) exhibits similar performance to Base-GT for GCN as it does not translate the format during FWP.
However, PyG performs poorly on NGCF, which is even worse than DGL by 15.7\%.
The reason for this performance degradation is memory bloat caused by Sparse2Dense, as analyzed in Figure \ref{fig:eval_breakdown_light}.
The memory bloat not only amplifies the GPU memory usage, but also wastes time by increasing the slow global memory access.
GNNAdvisor also suffers from memory bloat because it lacks a mechanism to compute edge weighting. As a result, users must rely on DL operations to implement edge weighting (DL-approach).
Moreover, GNNAdvisor performs slightly worse then PyG/Base-GT in GCN by 11.1\% due to its synchronization overhead.
It partitions neighbors into multiple neighbor groups and allocates them to different SMs, which makes multiple SMs updating the same output vector of a dst, thereby requiring synchronization.
Even though this mechanism mitigates the load imbalance across SMs in GNNAdvisor's target scenario (training GNN with a full graph), it offers little advantage in sampling-based training because the sampled graph is already well-balanced (see \textsection \ref{subsec:frontend}).
Note that GNN frameworks without sampling cannot handle graphs larger than the GPU memory, and therfore have limited scalability.

Compared to DGL and PyG, Base-GT shows 1.5$\times$ and 1.3$\times$ shorter GCN/NGCF training latency across all the workloads.
This performance enhancement mainly stems from NAPA's destination-centric, feature-wise scheduling, which removes the format translation and memory/cache bloats, as shown in Figure \ref{fig:eval_breakdown_light}.
Since Base-GT reduces the cache bloat, it further shortens the latency of edge weighting, compared to DGL and PyG by 2.5$\times$ and 1.8$\times$, respectively.
Figure \ref{fig:eval_gpu_resource} also shows NAPA's impact.
As shown in Figure \ref{fig:eval_membloat}, NAPA's scheduling reduces the memory footprint of FWP/BWP by 81.8\%, on average, by eliminating data copy for sparse-to-dense data conversion.
In addition, it reduces the amount of data loaded in to cache by 44.8\%, on average (Figure \ref{fig:eval_cachebloat}), by allocating dst's feature element to be processed within the same SM.
Thanks to the allocation, GPU also GPU reuses dst's features during edge weighting, thereby reducing global memory accesses.
In addition, in Figure \ref{fig:eval_inference_light}, the GCN and NGCF training latency of our Dynamic-GT outperforms even Base-GT by 47.7\% and 74.2\%, respectively.
This is because Dynamic-GT rearranges the kernel to follow the best execution order, reducing the dimension of embeddings by 4.1$\times$, on average.
Figure \ref{fig:eval_dkpanalysis} shows DKP's impact in other perspectives for the two representative workloads.
We measured the FLOPs and number of global memory accesses and normalized the result by that of Dynamic-GT.
As we can see in the figure, Dynamic-GT reduces the FLOPs by 5.4$\times$ as well as global memory access by 1.4$\times$, on average.

\noindent \textbf{Heavy feature graphs.}
Figure \ref{fig:eval_inference_heavy} shows the performance on heavy feature graphs.
Similar to light feature graph processing, Base-GT exhibits 1.3$\times$ faster performance than DGL and PyG, but the performance gap between DGL and Base-GT is reduced.
This is because the amount of computation for aggregation and combination severely increased owing to many edges and long features;
it makes the format translation overhead relatively small (Figure \ref{fig:eval_breakdown_heavy}).
Note that PyG and GNNAdvisor always perform worse than DGL when processing the graphs via NGCF.
Figure \ref{fig:eval_breakdown_heavy} shows that PyG and GNNAdvisor's NGCF spends a significant amount of time handling Sparse2Dense using the existing DL framework's primitives (32.3\% of the total NGCF latency).
For the worst case, \texttt{livejournal}, both frameworks are unable to process the NGCF training due to out-of-memory (Sparse2Dense).
This is because, \texttt{livejournal} has a large (4.3K) feature dimension and the number of edges is also greater than that of all other heavy-feature graphs (392.7K), as shown in Table \ref{tbl:dataset}.

In contrast, Dynamic-GT further reduces the GCN and NGCF training latency of Base-GT by 31.0\% and 11.4\%, respectively.
Similar to light feature graphs, this is because Dynamic-GT performs combination first if it brings benefits to reduce the number of embeddings to compute.
Note that the DKP performance gain is more promising for GCN than NGCF.
This is because the edge weighting that occupies 35.1\% of the execution time in NGCF, is hard to get benefit from kernel scheduling.

\subsection{End-to-End Performance Enhancement. \label{subsec:eval_e2e}}
Figure \ref{fig:e2e_exec_time} analyses the end-to-end latency (including the preprocessing). % beginning from the sampling to the end of GNN computing,
In our preliminary evaluations, we observed that PyG is severely slower than DGL and GraphTensor ($>$5$\times$) due to its single-threaded architecture for sampling.
We thus modify PyG to support multi-threaded preprocessing and compare it with other frameworks. %as the representative of DL-approach.
Specifically, we introduced a multi-thread pool where individual threads can simultaneously process different destination vertices.
In this evaluation, we exclude GNNAdvisor, since the original work does not support preprocessing.
Instead, we use another state-of-the-art preprocessing approach, SALIENT \cite{kaler2022accelerating}.
SALIENT allows for overlapping the execution times between preprocessing (including transfer) and the FWP/BWP in GPU while utilizing pinned memory for fast data transfer between CPU and GPU.
To ensure fair comparison, we normalized the results to the latency of Dynamic-GT.

As shown in Figure \ref{fig:e2e_exec_time}, DGL and Dynamic-GT perform better than PyG by 7.4\%. because they overlap the execution times between sampling/embedding lookup and the FWP/BWP in GPU.
However, the execution time of FWP/BWP is shorter than that of preprocessing, making the overlap less effective.
In addition, they all process neighbor sampling, reindexing, embedding lookup, and data transfers in serial order, which limits performance.
In contrast, SALIENT reduces the end-to-end latency by 19.7\% and 51.1\% for light and heavy feature graphs, respectively.
This is mainly due to their fast data transfer; SALIENT collects feature vectors in the pinned memory to avoid unnecessary data copy while transfering data from host memory to GPU memory.
Prepro-GT can further reduce the end-to-end latency by 1.7$\times$, on average.
This is because the service-wide tensor scheduler parallelizes the preprocessing execution by relaxing the dependency chain with smaller subtasks (and locks on the hash table).
The benefits of Prepro-GT over Dynamic-GT for light feature graphs are smaller compared to heavy feature graphs at some extent, which will be explained shortly.

Figures \ref{fig:eval_prepro_timeline} shows how our service-wide tensor scheduler can reduce the preprocessing times over two representative workloads. %(\texttt{products}/\texttt{wiki-talk}).
The y-axis is the number of nodes preprocessed, normalized by the total number of nodes in the sampled graph, and the x-axis is the accumulated time.
Even though Prepro-GT's completion time of sampling and reindexing is longer than Dynamic-GT's ones, embedding lookup and data transfers of Prepro-GT are completed earlier than those of Dynamic-GT by 14.9\% and 48.5\%, respectively.
As a result, it can shorten the preprocessing latency by 48.5\%, on average.
The reason why sampling and reindexing are slower than what the backend schedules the tensors is that Dynamic-GT runs each of them by fully utilizing all the cores of the host CPU,
whereas Prepro-GT shares the cores across other subtasks such as embedding lookup in parallel.
Note that the reason why light feature graphs exhibit a smaller benefit than heavy feature graphs is that the amount of data handled by embedding lookup and data transfers is smaller than light feature graphs.
As shown in Figure \ref{fig:eval_timeline_light}, the embedding lookup and data transfers cannot be started before the completion of sampling (orange arrow).
As a result, the preprocessing time of light feature graphs is bounded by the completion time of sampling.

%-------------------------------------------------------------------------------
\section{Related Work}
%-------------------------------------------------------------------------------
\label{sec:relatedwork}
% \markerDown[6]{3}{Minor}

Very recently, several studies explored new programming models and framework designs to make GNN processing efficient and accelerated.
As discussed in \cref{subsec:emerging_gnn_frameworks}, most of these works leverage a few successful public-frameworks, DGL and PyG, each representing DL-approach and Graph-approach in this paper.
We analyze diverse frameworks stemming from the two approaches and summarize their differences, including our GraphTensor in Table \ref{tbl:prior_gnn_frameworks}.
In particular, for each performance problems at the top of the table, the table shows if the frameworks have the problem (\checkyes) or not (\checkno).

DL-approach, such as GNNAdvisor \cite{osdi21gnnadvisor}, NeuGraph \cite{atc19neugraph}, and FlexGraph \cite{wang2021flexgraph}, suffer from the memory/cache bloats issues with sparse-to-dense data conversion overhead.
Specifically, GNNAdvisor
%\needtochange{offers aggregation methods that harmonize well with the existing DL operations. It also}
optimizes memory patterns and balances input graph data across different SM cores.
However, it has no mechanism to compute edge weighting, which cannot cover diverse GNN models. %and exhibit the memory/cache bloat issues.
NeuGraph and FlexGraph offer new programming methods similar to PyG by using scatter/gather primitives which require sparse-to-dense data conversion.
%Stellar and Spektral leverage TensorFlow and its default edge array tensor (COO), making users to implement edge weighting manually.

\begin{figure}[b]
  \centering
  \small
  \setlength{\tabcolsep}{6.3pt}
  \renewcommand{\arraystretch}{.95}
  \vspace{-18pt}
  \resizebox*{\linewidth}{!}{%
      \begin{tabular}{|cl|ccccc|}
          \hline
                                                                            &
          \multicolumn{1}{c|}{\multirow{3}{*}[0.15em]{\textbf{Frameworks}}} &
          \textbf{Initial}                                                  &
          \textbf{GPU}                                                      &
          \textbf{Format}                                                   &
          \textbf{GPU}                                                      &
          \textbf{Prepro-}                                                                                                                                                                                                                                                                                                                                           \\[-1.5pt]
                                                                            &                                   & \textbf{graph}                            & \textbf{memory} & \textbf{trans-}                        & \textbf{cache}                                                        & \textbf{cessing}                     \\[-1.9pt]
          %\cline{5-6}

                                                                            &                                   & \textbf{format}                           & \textbf{ bloat} &  \textbf{lation} & \textbf{ bloat}  & \textbf{overhead} \\\hline
          \multirow{4}{*}{\rotatebox[origin=c]{90}{\textbf{DL.}}}
                                                                            & PyG\cite{iclr19pyg}               & \textbf{\color{mygreen}\texttt{CSR}}      & \checkyes       & \checkno                                                             & \checkyes        & \checkyes         \\
                                                                            & NeuGraph\cite{atc19neugraph}      & \textbf{\color{mygreen}\texttt{CSR}}      & \checkyes       & \checkno                                                             & \checkyes        & \checkyes         \\
                                                                            & GNNAdvisor\cite{osdi21gnnadvisor} & \textbf{\color{mygreen}\texttt{CSR}}      & \checkyes       & \checkno                                                             & \checkyes        & \checkyes         \\
                                                                            & FlexGraph\cite{wang2021flexgraph} & \textbf{\color{mygreen}\texttt{CSR}}      & \checkyes       & \checkno                                                             & \checkyes        & \checkyes         \\\hline
          \multirow{4}{*}{\rotatebox[origin=c]{90}{\textbf{Graph.}}}
                                                                            & DGL\cite{wang2019dgl}             & \textbf{\texttt{COO}}                     & \checkno        & \checkyes                                                             & \checkyes        & \checkneutral     \\
                                                                            & FeatGraph\cite{hu2020featgraph}   & \textbf{\texttt{COO}}                     & \checkno        & \checkyes                                                             & \checkyes        & \checkneutral     \\
                                                                            & ROC\cite{MLSYS2020_gnnwithroc}    & \textbf{\color{mygreen}\texttt{CSR}}      & \checkno        & \checkyes                                                             & \checkyes        & \checkyes         \\
                                                                            & G3\cite{husong2020g3}             & \textbf{\texttt{COO}}                     & \checkno        & \checkyes                                                             & \checkyes        & \checkyes         \\\hline
                                                                            \rowcolor{gray!10}                                                & \textbf{GraphTensor}              & \textbf{\color{mygreen}\texttt{CSR}} & \checkno        & \checkno                                                              & \checkno         & \checkno          \\
                                                                            % ROC\cite{MLSYS2020_gnnwithroc}                                & \checkyes      & $-$            & \checkno  & \checkno           & --                  \\
          \hline
      \end{tabular}
      % parallel preprocessing, depenedency aware preprocessing scheduling, parallel preprocessing scheduling, preprocessing kernel scheduling, preprocessing overhead
  }
  \vspace{-5pt}
  \captionof{table}{Comparison across various GNN frameworks.}
  \vspace{-5pt}
  \label{tbl:prior_gnn_frameworks}
\end{figure}

% Specifically, GNNAdvisor optimizes memory patterns and balances input graph data such as a graph degree, but has no mechanism to compute edge weighting.
% Since these frameworks leverage PyTorch (PyG), it can exhibit the memory/cache bloat issues when the frameworks implement edge weighting primitives.
% NeuGraph and FlexGraph offer new programming methods basically similar to pull and apply of PyG, which induces the memory/cache bloats and thread synchronization issues on their (potential) implementation for edge weighting.
% Lastly, Stellar and Spektral leverage TensorFlow and its basic edge array tensor (COO), which makes users to manually implement edge weighting by having the sparse-to-dense data conversion just like other DL-approaches.

Graph-approach such as FeatGraph \cite{hu2020featgraph}, ROC \cite{MLSYS2020_gnnwithroc} can address the memory bloat issue, but they yet introduce performance degradation imposed by edge-wise kernel scheduling (e.g., cache bloat, format translation).
%FeatGraph reuses a machine learning compiler framework \cite{chen2018tvm} to support SpMM and SDDMM interfaces while ROC resembles a multi-GPU graph processing framework \cite{jia2017distributed}, to support balanced SpMM and SDDMM.
FeatGraph reuses a machine learning compiler framework to support SpMM/SDDMM interfaces while ROC resembles a multi-GPU graph processing framework, to support balanced SpMM and SDDMM.
However, they perform edge-wise kernel scheduling, not considering destination-centric graph traversing and (feature-wise) kernel scheduling. Therefore, they suffer from the cache bloat.
Note that ROC is the only GNN framework adopting CSR, but they use CSR for load balancing across GPUs, not thread scheduling. Therefore, it still needs to perform format translation (CSR to COO) during SDDMM. %\suggestion{In addition, ROC can perform SpMM without format translation by adopting CSR, it still needs to translate CSR into COO to perform SDDMM.}
% G3 also provides GNN APIs built on graph processing framework.
%\suggestion{추가로, ROC가 초기 포맷으로 CSR을 활용하기에 SpMM을 translation 오버헤드 없이 수행할 수 있으나, SDDMM을 위해서 CSR을 COO로 변환하는 과정이 필요하기에 translation 오버헤드가 사라지지 않음}
%Similarly, G3 provides GNN APIs built on graph processing framework \cite{wang2016gunrock}, which schedules all threads in an edge-wise manner, thereby suffering from the cache bloat.
%Similarly, G3 provides GNN APIs built on graph processing framework that schedules all threads in an edge-wise manner, thereby suffering from the cache bloat.

%There are a few studies that consider reducing GNN preprocessing overhead.

% \markerDown[1]{3}{Minor}

There are a few studies trying to reduce GNN preprocessing overhead.
DGL and Featgraph use multiple threads for preprocessing, but yet have limited performance as shown in \textsection \ref{subsec:eval_e2e} ($\checkneutral$ in Table \ref{tbl:prior_gnn_frameworks}).
FlexGraph \cite{wang2021flexgraph} and NextDoor \cite{jangda2020nextdoor} load the entire graph to multiple GPUs and accelerate neighbor sampling.
These studies assume that all graph data can be accommodated by GPU memory, thus cannot handle large graph datasets that we target to process.
PaGraph \cite{lin2020pagraph} caches frequently referred embeddings in GPU's internal DRAM, thereby reducing data transfer latency.
The work unfortunately requires high locality on sampled data, and its effectiveness significantly varies on the input datasets and user behaviors.

Note that none of these frameworks is aware of the dimensionality reduction and rearranges the kernel execution sequence.
There is an approach aiming to integrate aggregation and edge weighting together, called FusedMM \cite{rahman2020fusedmm}.
However, it works on a CPU-only system and has no consideration of taking GPU's massive computing with parallelism.

% AGL\cite{zhang2020agl} : GNN이 대규모 그래프를 학습할 수 있도록 하는 파라미터 서버 방식의 GNN 분산 학습 시스템을 제안.
% 분산시스템에 저장된 그래프에서 Map-reduce를 이용하여 샘플링을 수행하는 방법을 제안하며, CPU 를 이용하여 GNN을 학습하는 경우를 다루어 우리 work과 orthogonal함.
% 우리 work과 integrate 하여 GPU 기반 GNN 분산 학습 프레임워크를 구현할 수 있을 것으로 예상

%-------------------------------------------------------------------------------
\section{Conclusion}
%-------------------------------------------------------------------------------
\label{sec:conclusion}
In this paper, we introduce GraphTensor, a novel open-source framework that supports efficient parallel neural network processing on large graphs.
GraphTensor offers a set of easy-to-use programming primitives that appreciate both graph and neural network execution characteristics from start to finish.
Our evaluation shows that GraphTensor outperforms modern GNN frameworks (DGL and PyG) by 1.4$\times$ in training and 2.4$\times$ across diverse large-scale graph workloads.

\section*{Acknowledgment}
We thank the anonymous reviewers for their constructive feedback.
This work is mainly supported by Samsung (SRFC-IT2101-04). This work is also supported in part by IITP's 2021-0-00524 \& 2022-0-00117, and NRF's 2021R1A2C4001773.
GraphTensor is protected by one or more patents.
Myoungsoo Jung is the corresponding author (mj@camelab.org).

\bibliographystyle{abbrev_first}
\bibliography{ref}

\end{document}